\documentclass[12pt,oneside,onecolumn,floatfix]{article}

\usepackage{epsfig}
\usepackage{a4wide}
\usepackage[top=30pt,bottom=30pt,left=48pt,right=46pt]{geometry}
\usepackage{slashed, enumitem}
\usepackage[utf8]{inputenc}
\usepackage{jcappub}

\usepackage{float,color}
\usepackage{subfigure}
\usepackage{multirow}
\usepackage{slashed}
\usepackage{placeins}
\usepackage{wasysym} 
\usepackage{mathrsfs} 
\usepackage{caption}
\usepackage{amsmath}
\usepackage{amssymb}
\usepackage{graphicx}
\usepackage{slashed}
\usepackage{multirow}
\usepackage{placeins}
\usepackage[dvipsnames]{xcolor}
\usepackage{epstopdf}
\usepackage{soul}
\usepackage{tikz}
\usepackage[capitalise, english]{cleveref}
\usepackage{siunitx}
\usepackage{xspace}
\usetikzlibrary{trees}
\usetikzlibrary{decorations.pathmorphing}
\usetikzlibrary{decorations.markings}
\usepackage[colorlinks=true,citecolor=blue,linkcolor=blue]{hyperref}

\usepackage{mhchem}

\usepackage{subfigure}

\usepackage{upgreek}
\usepackage{mathtools}
\usepackage{setspace, slashed, gensymb, hhline, booktabs}
\usepackage{wrapfig}
\usepackage{makecell}

\usepackage{pifont}% http://ctan.org/pkg/pifont
\newcommand{\cmark}{{\footnotesize \ding{51}} }%
\newcommand{\xmark}{{\footnotesize \ding{55}}}%

\definecolor{viridian}{rgb}{0.25, 0.51, 0.43}
\definecolor{mediumseagreen}{rgb}{0.24, 0.7, 0.44}
\definecolor{otterbrown}{rgb}{0.4, 0.26, 0.13}
\definecolor{saddlebrown}{rgb}{0.55, 0.27, 0.07}
\definecolor{americanrose}{rgb}{1.0, 0.01, 0.24}
\definecolor{ao}{rgb}{0.0, 0.0, 1.0}

\newcommand\myshade{80}
\colorlet{mylinkcolor}{violet}
\colorlet{mycitecolor}{red}
\colorlet{myurlcolor}{ao}

\newcommand{\NC}{\text{NC}}
\newcommand{\CC}{\text{CC}}

\hypersetup{
	linkcolor  = mylinkcolor!\myshade!black,
	citecolor  = mycitecolor!\myshade!black,
	urlcolor   = myurlcolor!\myshade!black,
	colorlinks = true
}

\usepackage{tikz,xcolor,hyperref}

\definecolor{lime}{HTML}{A6CE39}
\DeclareRobustCommand{\orcidicon}{\hspace{-3mm}
	\begin{tikzpicture}
		\draw[lime, fill=lime] (0,0) 
		circle [radius=0.16] 
		node[white] {\hspace{0.1mm}{\fontfamily{qag}\selectfont \tiny ID}};
		\draw[white, fill=white] (-0.07,0.1) 
		circle [radius=0.01];
	\end{tikzpicture}
	\hspace{-5mm}
}

\foreach \x in {A, ..., Z}{\expandafter\xdef\csname 
orcid\x\endcsname{\noexpand\href{https://orcid.org/\csname orcidauthor\x\endcsname}
		{\noexpand\orcidicon}}
}

\title{A deuterated liquid scintillator for supernova neutrino detection}

\author[a\hspace{-5pt}]{Bhavesh Chauhan\orcidA{}}
\emailAdd{bhavesh@theory.tifr.res.in}

\author[a\hspace{-5pt}]{, Basudeb Dasgupta\orcidB{}}
\emailAdd{bdasgupta@theory.tifr.res.in}

\author[b]{,\\ and Vivek Datar\orcidC{}}
\emailAdd{vivek.datar@gmail.com}

\affiliation[a\hspace{1pt}]{Tata Institute of Fundamental Research,\\ Homi Bhabha Road, Mumbai 
	400005, 
	India}
\affiliation[b\hspace{1pt}]{Raja Ramanna Fellow, The Institute of Mathematical Sciences, \\ CIT 
	Campus, 
	Taramani, Chennai 600113, India}

\abstract{
	For the next galactic supernova, operational neutrino telescopes will measure the 
	neutrino flux several hours before their optical counterparts. Existing detectors, relying mostly on 
	charged current interactions, are mostly sensitive to $\bar{\nu}_e$ and to a lesser extent to 
	$\nu_e$. In order to measure the flux of other flavors 
	($\nu_{\mu},\bar{\nu}_{\mu},\nu_{\tau},\text{and}~\bar{\nu}_{\tau}$), we need to observe their 
	neutral current interactions with the detector. Such a measurement is not only crucial for overall 
	normalization of the supernova neutrino flux but also for understanding the intricate neutrino 
	oscillation physics.  A deuterium based detector will be sensitive to all neutrino flavors. In this 
	paper, we propose a 1\,kton deuterated liquid scintillator (DLS) based detector that will see 
	about 435 neutral current events and 170 (108) charged current $\nu_e$  ($\bar{\nu}_e$) 
	events from a fiducial supernova at a distance of 10 kpc from Earth. We explore the possibility of 
	extracting spectral information from the neutral current channel $ 
	\overset{\scriptscriptstyle(-)}{\nu} d \rightarrow  \overset{\scriptscriptstyle(-)}{\nu}np$ by 
	measuring the quenched kinetic energy of the proton in the final state, where the neutron in the 
	final state is tagged and used to reduce backgrounds. We also discuss the secondary 
	interactions of the recoil neutrons in the detector.
}

\subheader{TIFR/TH/21-5}


\begin{document}
	
	\maketitle

	\section{Introduction}
	
	On average, a supernova occurs every second in our observable universe. The latest theoretical 
	estimates for the rate of core-collapse supernova in Milky Way is approximately 1-2 per 100 
	years 
	\cite{Reed:2005en, Rozwadowska:2021lll}. The last nearby supernova was seen on 23 February 
	1987 when a $\sim20M_{\odot}$ blue supergiant exploded in the Large Magellanic Cloud 
	approximately 50 kpc away. Neutrino detectors that were operational at the time, e.g., 
	Kamiokande-II \cite{Hirata:1987hu, Hirata:1988ad}, IMB \cite{Bionta:1987qt}, and Baksan 
	\cite{Alekseev:1987ej}, observed a total of 24 neutrino events, mostly from the electron 
	anti-neutrinos, a few hours before the optical signal. Despite the low statistics, these neutrinos 
	from SN1987 have helped in establishing a baseline model of supernova explosion and provided 
	a 
	wealth of constraints on new physics scenarios \cite{Ellis:1987pk, Cullen_1999, Raffelt:1987yt, 
		Turner:1987by, Mayle:1987as, Choi:1989hi, Gandhi:1990bq, Arafune:1987ua, Arnett:1987iz, 
		Kolb:1987dda, Kolb:1987qy, HariDass:1987ss}. It is now widely held that 
	observation of supernova neutrinos, from the galaxy or even farther away, holds great promise 
	for 
	fundamental physics and astrophysics~\cite{Mirizzi:2015eza, Horiuchi:2017sku}.\\
	
	Although we now have significantly larger neutrino detectors \cite{Scholberg:2012id}, there 
	remain critical deficiencies in our ability to detect supernova neutrinos. Our best instruments for 
	the purpose, such as Super-Kamiokande \cite{Ikeda:2007sa}, LVD \cite{Aglietta:1992dy, 
	Agafonova:2007hn}, HALO \cite{Duba:2008zz}, and \mbox{IceCube}\footnote{Even though the 
	primary 
		goal of IceCube is to observe TeV neutrinos, it too can detect supernova neutrinos through 
		the 
		correlated increase in the dark noise rate of its photomultiplier tubes 
		\cite{Halzen:1995ex}.}\,\cite{Abbasi:2011ss}
	are mostly sensitive to only the $\bar\nu_e$ flavor, as they depend on the inverse beta 
	decay ($\overline{\nu}_e + p \rightarrow n + e^+$). As a result, the community has set its hopes 
	on 
	future experiments such as DUNE \cite{Abi:2020wmh, Abi:2020evt}, JUNO \cite{An:2015jdp,  
		Djurcic:2015vqa, Abusleme:2021zrw}, and THEIA~\cite{Askins:2019oqj} to provide sensitivity 
		to 
	$\nu_e$ and the non-electron flavors, respectively. Hyper-Kamiokande \cite{Abe:2011ts} will 
	improve the $\bar\nu_e$ statistics by several orders of magnitude. In this paper, inspired by the 
	success of the previous large-scale deuterium based detector SNO~\cite{Jelley:2009zz}, we 
	propose a kton scale deuterated liquid scintillator (DLS) based detector, doped with Gadolinium 
	(Gd), and instrumented with PMTs, which can detect \emph{all the 
		flavors} of supernova neutrinos with spectral information, typically with reduced backgrounds.\\
	
	The main challenge in deuterium based detectors is the availability of deuterium. India is the one 
	of 
	the world's largest producers of heavy water, and has the capability to produce a variety of 
	deuterated compounds including deuterated hydrocarbons.\footnote{See for example at 
		\url{https://www.hwb.gov.in/}} It is therefore our understanding that a kton-scale deuterated 
	scintillator detector is realistically achievable. For concreteness, we assume a chemical 
	composition of the form $\text{C}_n \text{D}_{2n}$, mimicking an ordinary scintillator. In 
	practice, 
	significant research is required into identifying a suitable candidate DLS. Much of what we 
	discuss 
	here could also work with 
	a heavy water detector with water soluble ordinary scintillator. We also note that a DLS has 
	previously been used to study neutron capture, albeit with a much smaller detector 
	\cite{159660}. 
	To the best of our knowledge, this is the first discussion about using them for neutrino physics.\\
	
	Neutrino detection in a DLS will be dominantly via dissociation of the 
	deuteron. This proceeds via neutral current (NC) as $\nu+d\to \nu+p+n$ or $\overline\nu+d\to 
	\overline\nu+p+n$ for all 
	three flavors, and via charged current 
	(CC) as $\nu_e+d\to e^-+p+p$ or 
	$\overline\nu_e+d\to e^+ +n+n$ for the electron flavor. While the outgoing $\nu$ or $\bar{\nu}$ 
	cannot be detected, all the other 
	particles have characteristic signatures. For typical supernova neutrinos with $E_\nu\sim 
	(5-50)$\,MeV, the 
	protons have recoil energies $\sim(0.1-1)$\,MeV and are highly non-relativistic. These lead to 
	quenched scintillation signals arising from their high ionization loss $\sim 
	(10^3-10^2)$\,MeV\,cm$^{-1}$. 
	The final state neutrons thermalize due to elastic collisions with the neighboring nuclei and get 
	captured, emitting detectable photons.\footnote{We note that a thermal neutron travels 
		$\sim170$\,cm in heavy water before capture\,\cite{doi:10.1139/cjr47a-015}, compared to a 
		few cm 
		in ordinary water \cite{DeJuren}. The diffusion length of neutrons in DLS will be much smaller 
		if 
		$n$-capture dopants, such as Gd, are present. For a more detailed discussion see 
		section \ref{sec:Secondary}.} The 
	final-state electrons and positrons have 
	energies similar to the incoming neutrinos, thus are highly relativistic, and produce 
	Cherenkov radiation with $\sim 270$ photons\,cm$^{-1}$\,eV$^{-1}$ \cite{Zyla:2020zbs}.  This 
	Cherenkov light arrives earlier than scintillation but is likely to be absorbed and/scattered away 
	by 
	the scintillator.\\
	
	\begin{table}[!t]
		\centering
		\caption{\label{tab:comp} Comparison of final state and detection channels between ordinary 
		detectors
			and a deuterated scintillator detector. We have not shown some channels that would be 
			common to both, 
			e.g., $\nu-e$ and $\nu-C$ interactions. In the table, \emph{Spectrum} refers to the ability 
			of 
			the detector to reconstruct the incident neutrino spectrum from the channel, whereas 
			\emph{Tagging} refers to having more than one detectable final state particle that can be 
			used to tag signal events against backgrounds.}
		{\small
			\begin{tabular}{c c c c c}
				\toprule
				\multirow{2}{*}{Flavor} & \multicolumn{2}{c}{Ordinary Detector}  & 
				\multicolumn{2}{c}{Deuterated Detector} \\
				& Channel & Detector &Channel & Detector  \\
				\midrule
				\multirow{4}{*}{$\overset{\scriptscriptstyle(-)}{\nu}$} & 
				$\overset{\scriptscriptstyle(-)}{\nu} p 
				\to 
				\overset{\scriptscriptstyle(-)}{\nu} p $ & \makecell{JUNO, THEIA \\ \emph{Spectrum} 
					\cmark
					\\ \emph{Tagging}  \xmark} & 
				$\overset{\scriptscriptstyle(-)}{\nu}d\to \overset{\scriptscriptstyle(-)}{\nu}  
				~p~n$ & \makecell{SNO \\ \emph{Spectrum} \xmark \\ \emph{Tagging} \xmark}   \\[.75cm]
				
				& $\overset{\scriptscriptstyle(-)}{\nu} \text{ Pb} 
				\to 
				\overset{\scriptscriptstyle(-)}{\nu} \text{ Pb*} $ & \makecell{HALO \\ \emph{Spectrum} 
				\xmark 
					\\ 
					\emph{Tagging} \xmark}& $\overset{\scriptscriptstyle(-)}{\nu}d\to 
				\overset{\scriptscriptstyle(-)}{\nu}  
				~p~n$ &   \makecell{DLS \\ \emph{Spectrum} \cmark \\ \emph{Tagging} \cmark}  \\[.75 
				cm]

				\midrule
				\multirow{4}{*}{$\nu_e$} & $ \nu_e \text{ Ar} \to e^- \text{K*}$ & \makecell{ DUNE \\ 
					\emph{Spectrum} 
					\cmark
					\\ \emph{Tagging}  \xmark} & 
				$\nu_e~d \rightarrow e^- ~p~p$ &  \makecell{ SNO \\  \emph{Spectrum} \cmark \\ 
					\emph{Tagging} 
					\xmark}  \\[.75cm]
				
				& $ \nu_e \text{ Pb} 
				\to 
				\nu_e \text{ Bi*} $ & \makecell{ HALO \\  \emph{Spectrum} \xmark \\ 
					\emph{Tagging} \xmark} & $\nu_e~d \rightarrow e^- ~p~p$ & \makecell{DLS \\  
					\emph{Spectrum} 
					\cmark \\ \emph{Tagging}  {\color{gray} \cmark}}  \\[0.75cm]

				\midrule
				\multirow{4}{*}{$\bar\nu_e$} & $\overline{\nu}_e~p \rightarrow e^+  ~n$ & 
				\makecell{SuperK+Gd \\ \emph{Spectrum}
					\cmark \\ 
					\emph{Tagging} \cmark} & $\overline{\nu}_e  d \rightarrow e^+ ~n ~n$&   
				\makecell{SNO \\ \emph{Spectrum}  \cmark \\ 
					\emph{Tagging} \cmark } \\[0.75cm]
				& $\overline{\nu}_e~p \rightarrow e^+  ~n$ & \makecell{LVD, JUNO \\ \emph{Spectrum}  
					\cmark \\ 
					\emph{Tagging} \xmark} & $\overline{\nu}_e  d \rightarrow e^+ ~n ~n$&   
				\makecell{DLS \\ \emph{Spectrum} \cmark \\ 
					\emph{Tagging} \cmark } \\
				\bottomrule
			\end{tabular}
		}
	\end{table}
	
	In Table\,\ref{tab:comp}, we give a bird's eye view comparison of a DLS detector with other 
	prominent neutrino detectors.  For the non-electron flavors, $\nu_{\mu}, \bar{\nu}_{\mu}$, 
	$\nu_{\tau}, \text{and}~\bar{\nu}_{\tau}$, NC elastic scattering of a neutrino (or antineutrino) 
	with proton, 
	$\overset{\scriptscriptstyle(-)}{\nu} p \rightarrow 
	\overset{\scriptscriptstyle(-)}{\nu}p$~\cite{Beacom:2002hs}, can be 
	detected in ordinary scintillator detectors such as JUNO \cite{An:2015jdp,  Djurcic:2015vqa, 
		Abusleme:2021zrw} and water based scintillator such as THEIA \cite{Askins:2019oqj}. With 
		electrons, the 
	relevant cross section is smaller by a ratio $m_e/E_\nu$. Although the scintillation signal from 
	the 
	recoil proton is quenched, the one-to-one correspondence between proton kinetic energy and 
	quenched scintillation energy allows for reconstruction of the incident neutrino 
	energy~\cite{Dasgupta:2011wg}. In contrast, for a DLS, the NC interaction is observed through $ 
	\overset{\scriptscriptstyle(-)}{\nu} d \rightarrow  \overset{\scriptscriptstyle(-)}{\nu}np$. The 
	proton is 
	detected as for an ordinary scintillator. In addition, the final state neutron can be captured. This 
	allows for a possibility to tag the scintillation signal using the associated 
	neutron event and get rid of single-scintillation backgrounds. SNO, a Cherenkov detector, was 
	not 
	sensitive to the low-energy proton recoil and thus had no spectral information on these events. 
	Moving to CC events, ordinarily, $\bar{\nu}_e$ interacts via inverse beta decay $\overline{\nu}_e 
	+ 
	p \rightarrow e^+ + n$. The scintillation photons from positron and 
	the delayed photons from the neutron capture provide a clean signal. On the other hand, in a 
	DLS 
	the interaction proceeds through $\overline{\nu}_e + d \rightarrow e^+ + n + n$. The additional 
	neutron in the final state doubles the neutron capture efficiency and gives a unique and 
	essentially 
	background-free event signature. Ordinary scintillator 
	detectors can, in principle, observe $\nu_e$ through its interaction on Carbon. However, due to 
	large threshold and smaller cross section, the sensitivity is expected to be poor 
	\cite{Fukugita:1988hg}. Liquid Argon based detectors like DUNE \cite{Abi:2020wmh, 
	Abi:2020evt}, are hoped to 
	be more sensitive, but significant experimental work remains~\cite{Friedland:2018vry}. Several 
	other ideas have been investigated recently~\cite{Laha:2013hva, 
	Laha:2014yua,Nikrant:2017nya}. Detecting $\nu_e$ from a supernova explosion is therefore 
	challenging but important, 
	especially to study the neutronization burst \cite{Kachelriess:2004ds}. A DLS is sensitive to 
	$\nu_e$ through the CC 
	interaction $ \nu_e + d \rightarrow e^- + p + p$. In comparison with HALO \cite{Duba:2008zz}, 
	which is a dedicated supernova neutrino detector at SNOLAB, a DLS based detector can be built 
	to be significantly larger. Further, with lead (Pb) as target in HALO, the threshold for neutrino NC 
	and CC interaction is around 10 MeV, whereas for DLS the thresholds are 1.44 (4.03) MeV for 
	$\nu_e$ ($\bar{\nu}_e$) interacting  via CC and 2.22 MeV for NC interactions. Thus, a DLS 
	detector will not only observe larger number of events, but also lower energy neutrinos that 
	cannot 
	be detected by HALO.\\
	
	Radioactive backgrounds are a crucial concern. If we limit ourselves with backgrounds that 
	could 
	be relevant for galactic supernova neutrino detection, we focus on anything with an event rate 
	that 
	exceeds about 1 per second per kton. In ordinary scintillator detectors  the main background 
	comes from the beta decays of $^{14}{\rm C}$ that creates a ``wall'' of photons at 200 keV 
	and 
	below.  For a DLS, photons with energy larger than 2.2 MeV photodissociate the $d$ in the 
	detector material and pose a potential problem. Beta decays of $^{208}{\rm Tl}$ and 
	$^{214}{\rm 
		Bi}$ in the $^{238}{\rm U}$ and $^{238}{\rm Th}$ chains produce such photons. In SNO 
		great 
	care was taken to reduce this background to about a few hundred per year above 5 MeV. Lower 
	energy photons due to intrinsic and extrinsic radioactivity created a ``wall'' of photons below 5 
	MeV.  For galactic supernova neutrino detection in DLS, this low energy background could 
	obscure 
	the proton recoil signal. Clearly, one must reduce this background. For NC events the 
	neutrino-induced proton recoil is accompanied by a neutron capture signal. Using the spatial 
	correlation of the neutron capture and proton recoil may help isolate the proton recoil signal 
	away 
	from the background, at least for some fraction of the events. A dedicated study of possible 
	backgrounds and mitigation techniques is required.\\

	A DLS based detector therefore offers unique advantages, but is not without its challenges. One 
	could detect significant numbers of CC events from both $\nu_e$ and $\bar\nu_e$ and NC 
	events 
	from all flavors, with spectral information and reduced backgrounds within the same set up. 
	However, in addition to all the usual challenges of building a large scintillator detector with 
	extreme 
	radio-purity, one has the overhead of processing and handling the deuterated chemicals. These 
	challenges, while daunting, are not technologically forbidding. We believe the science-case is 
	promising to motivate further study.\\
	
	This paper is organized as follows. In Section\,\ref{methods}, we discuss the cross sections for 
	neutrino dissociation of deuteron at supernova neutrino energies, and outline our assumptions 
	for 
	supernova neutrino fluxes. In Section\,\ref{tot}, we estimate the total number of events seen in 1 
	kton of DLS for a galactic supernova. As several hundred events are predicted, in 
	Section\,\ref{spec} we explore the possibility to measure the spectrum of events for the four 
	channels. In Section\,\ref{sec:Secondary}, we discuss the secondary interactions of recoiling 
	neutrons and emphasize the need for adding Gd to the detector. We discuss the avenues for 
	further research and summarize our results in Section\,\ref{conc}. 
	
	\section{Methods}\label{methods}
	
	\subsection{Neutrino deuteron interaction cross section} \label{nudxs}
	
	The cross section for neutrino dissociation of deuteron can be calculated in two independent 
	ways. 
	First, using the Phenomenological Lagrangian Approach (PhLA) employed by Nakamura, Sato, 
	Gudkov, and Kubodera (NSGK) \cite{Nakamura:2000vp}. Second, in the framework of Pionless 
	Effective Field Theory ($\slashed{\pi}$EFT) valid for 
	$E_\nu \ll m_\pi$ \cite{Chen:1999tn} used by Butler, Chen, and Kong (BCK) 
	\cite{Butler:2000zp}. The total cross sections obtained by both methods are in excellent 
	agreement. A discussion about the two approaches can be found in Appendix \ref{revist}. In 
	Figure\,\ref{comp}, we see that the total cross sections obtained by NSGK and BCK are very 
	similar 
	up to neutrino energy of 50 MeV. Thus we can safely use $\slashed{\pi}$EFT for estimating the 
	supernova neutrino event rates. \\
	
	\begin{figure}[!h]
		\centering
		\includegraphics[width=7cm]{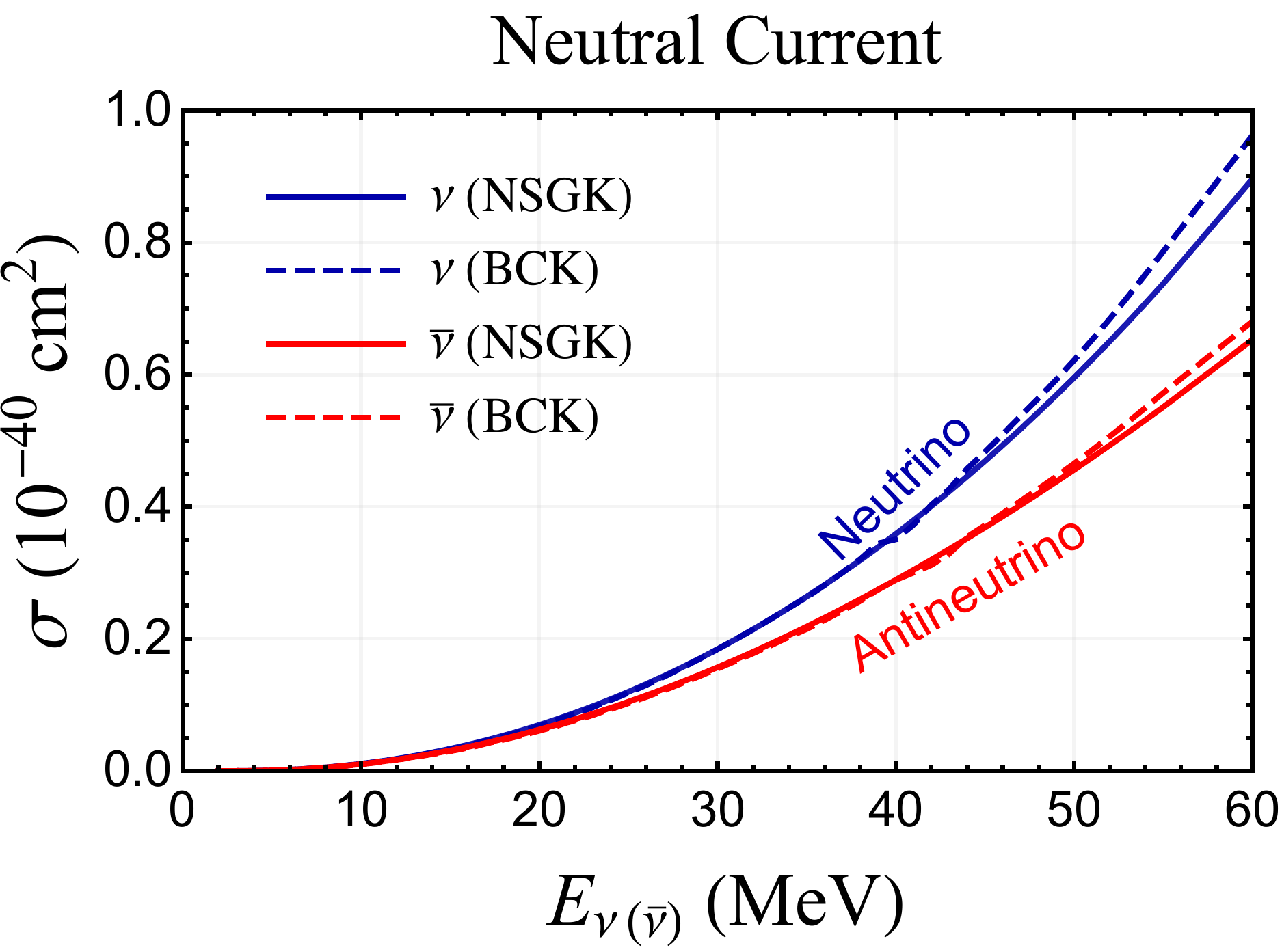}\hspace{0.5cm}
		\includegraphics[width=7cm]{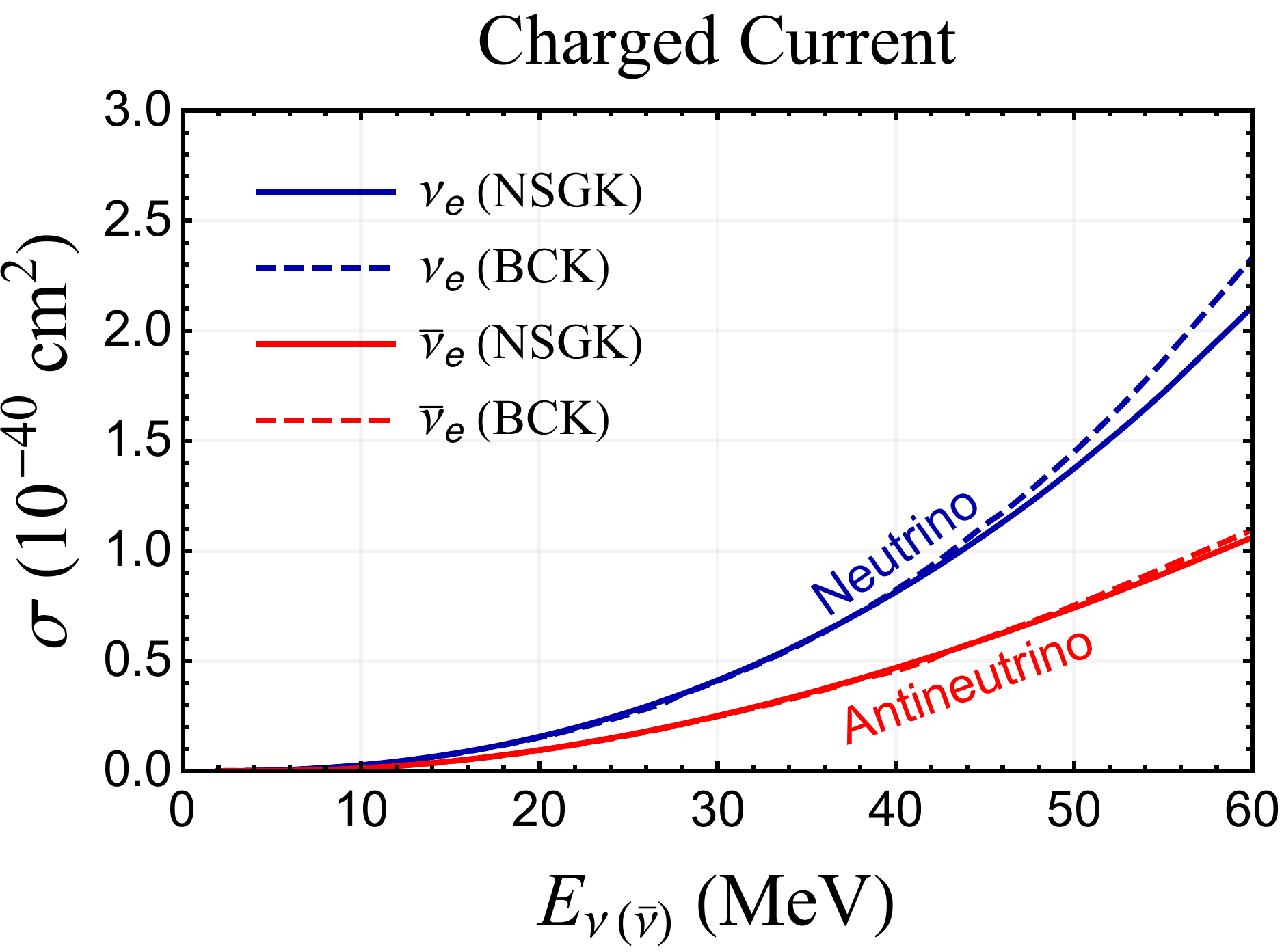}
		\caption{\label{comp} The cross section for dissociation of deuteron by neutrino (blue) and 
			antineutrino (red) as a function of incident neutrino or antineutrino energy is shown for 
			Neutral 
			Current (left) and Charged Current (right) interactions. 
			The cross section obtained using PhLA by NSGK \cite{Nakamura:2000vp} is shown by solid 
			lines and the one obtained using $\slashed{\pi}$EFT by BCK \cite{Butler:2000zp} is shown 
			by 
			dashed lines.}
	\end{figure}

	For this paper, we have used the tabulated numerical values from NSGK where total cross 
	sections 
	are required. However, for the differential cross sections, NGSK only provides plots for a few 
	energies, whereas BCK provides analytical expressions in terms of the outgoing lepton energy, 
	which is unmeasurable for the NC channel. We therefore recomputed all the differential cross 
	sections in terms of measurable energies, i.e., specifically for the proton in NC channel, using 
	the 
	method of BCK. As a cross check we compared the total cross section obtained from the above 
	to 
	the NGSK results. In doing so, we found anomalies that were traced back to a typo in 
	one of the expressions in BCK.\footnote{We thank J. W. Chen for sharing with us his original 
		calculations and pointing out the typo in the published version, following which we could 
		reproduce 
		their results.} After these corrections, the differential cross sections thus obtained and 
	integrated 
	over energy agree with the total cross sections in NGSK to better than a few per cent for 
	neutrino 
	energies up to 50 MeV. This is apparent from Fig.\,\ref{comp}. The differential cross sections, 
	thus 
	obtained (see Appendix \ref{revist} for details), are shown in Figure \ref{fig:diffxsNC} for NC and 
	in 
	Figure \ref{fig:diffxsCC} for CC 
	interactions. In Appendix~\ref{tabxsec}, we provide tabulated NC and CC differential cross 
	sections 
	in terms of the incoming neutrino energy and the \emph{measurable} outgoing proton or 
	electron/positron energy, respectively\footnote{Tabulated differential and total cross sections 
	are 
		also available at \url{https://github.com/bhvzchhn/NeutrinoDeuteron/}.}.\\
	
	\begin{figure}[!h]
		\centering
		\includegraphics[width=7cm]{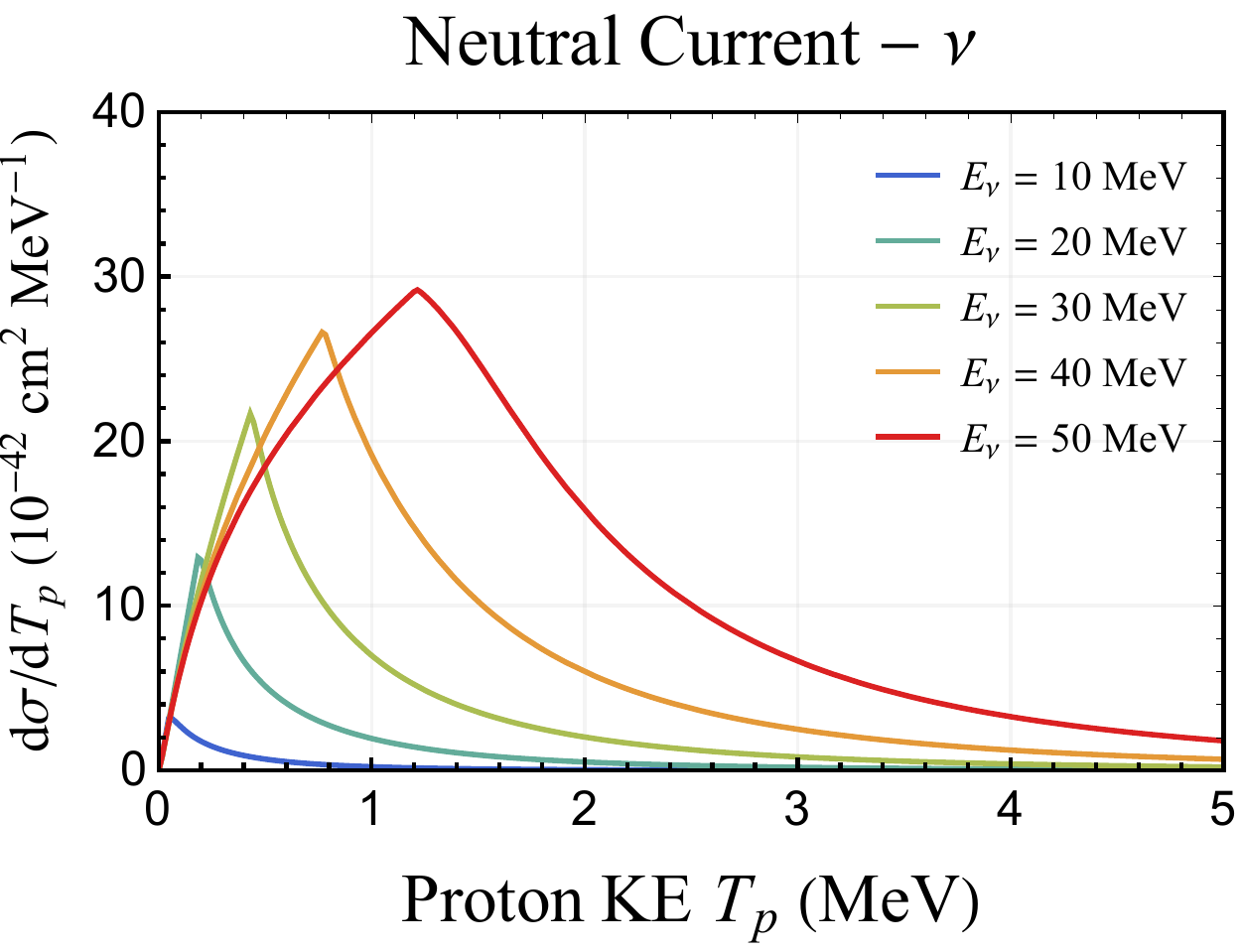} \hspace{0.5cm}
		\includegraphics[width=7cm]{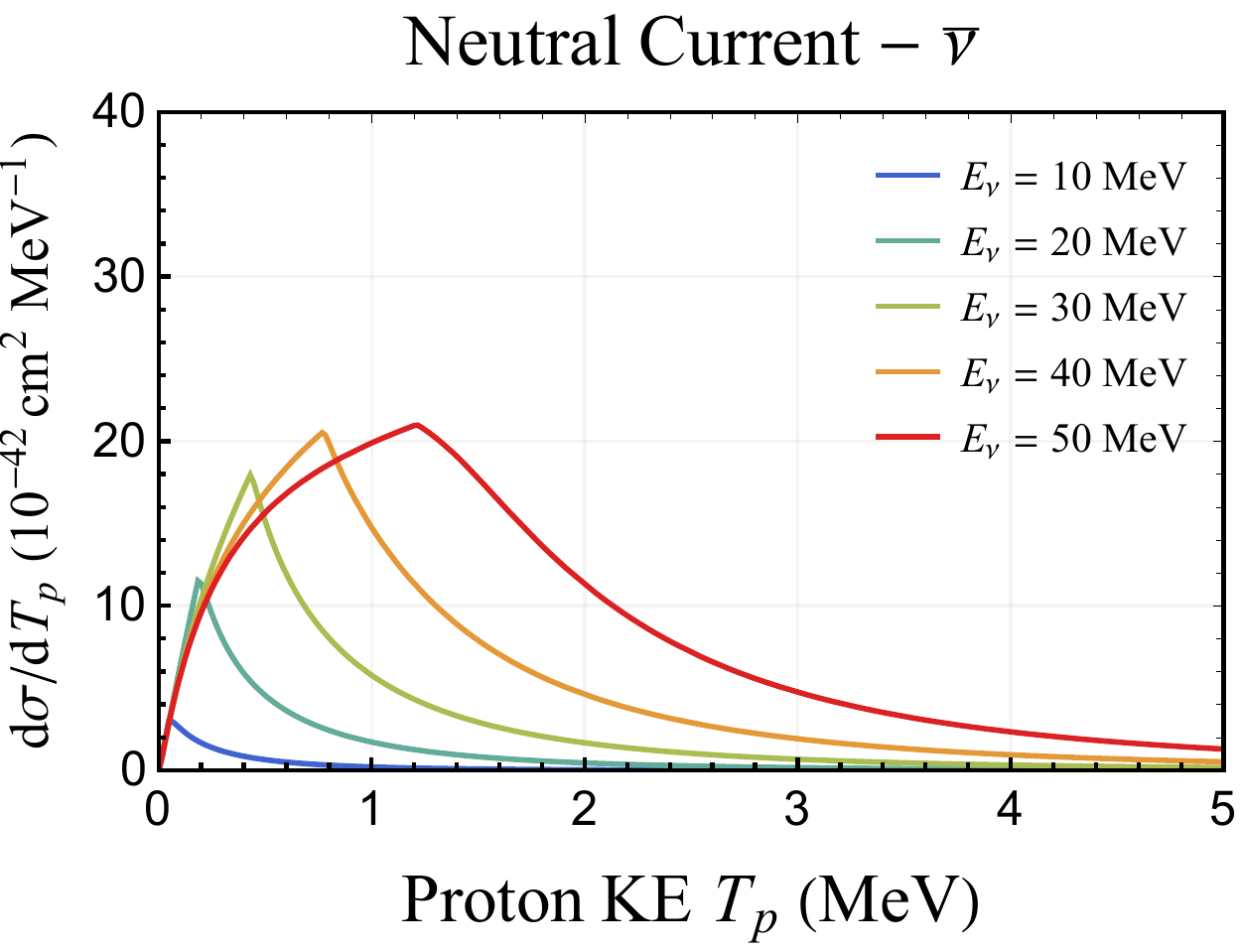}
		\caption{\label{fig:diffxsNC} The differential cross section for dissociation of deuteron by 
			neutrino (left) and antineutrino (right) for the Neutral Current interaction channel as a 
			function 
			of 
			kinetic energy of final state proton is shown for various choices incident neutrino or 
			antineutrino 
			energy. The difference between neutrino and antineutrino interaction strengths (arising 
			from 
			the 
			sign of $W_3$ term in Eq.\eqref{eq:diff}) is apparent. One can also infer that the proton 
			carries 
			only a fraction of incident energy, while the majority is carried away by the final state 
			neutrino 
			which is unmeasurable.}
	\end{figure}
	
	\begin{figure}[!h]
		\centering
		\includegraphics[width=7cm]{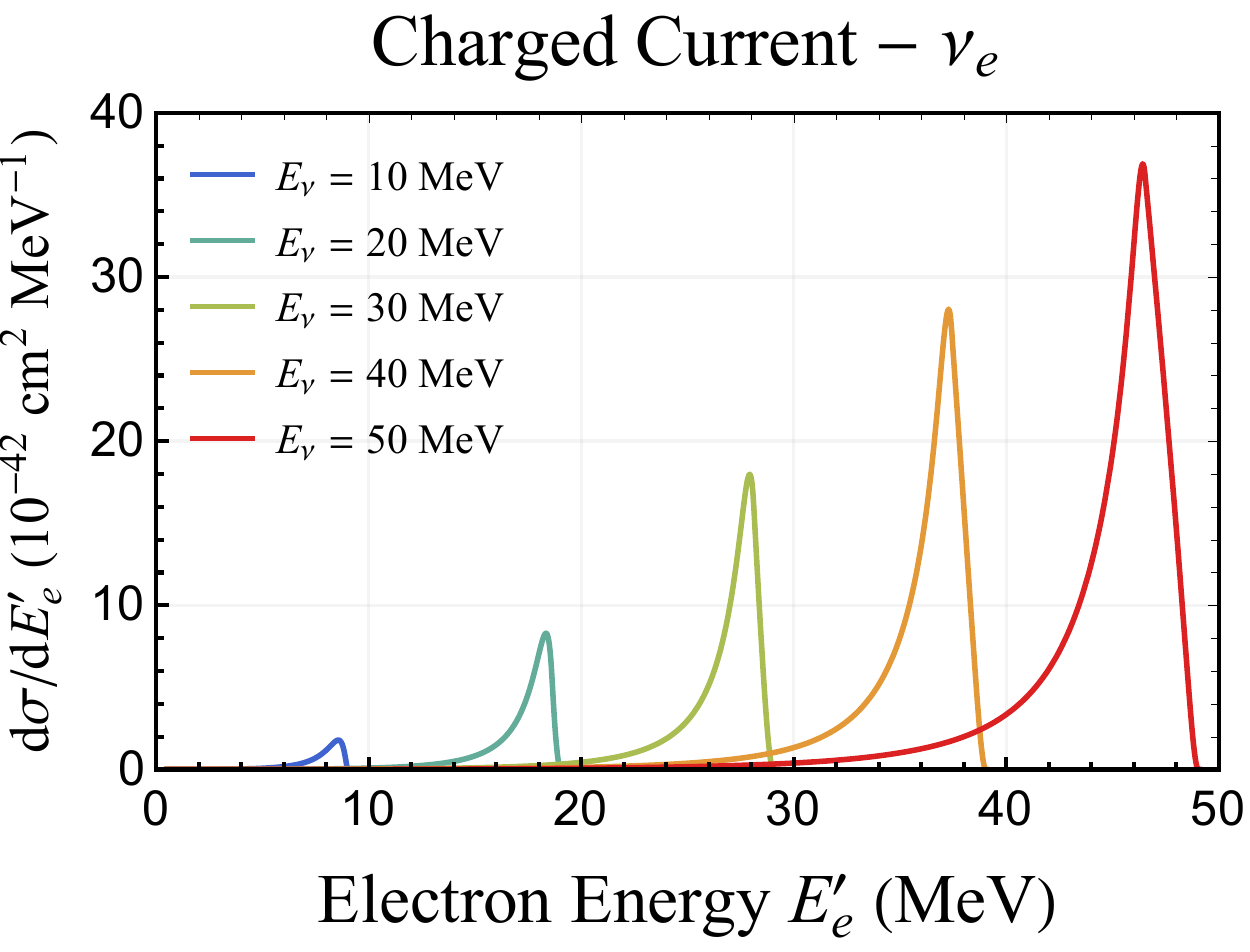} \hspace{0.5cm}
		\includegraphics[width=7cm]{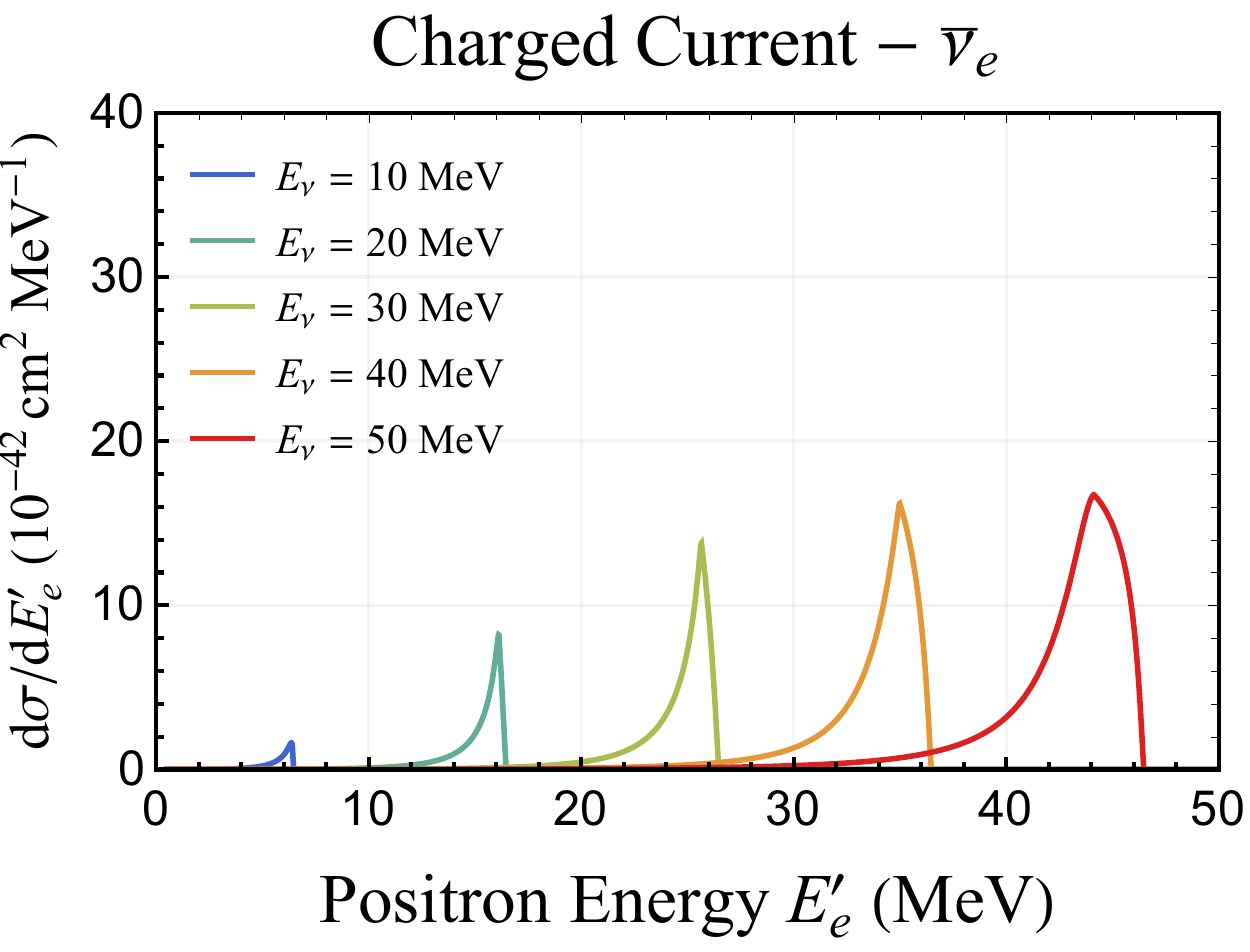}
		\caption{\label{fig:diffxsCC} The differential cross section for dissociation of deuteron by 
			neutrino (left) and antineutrino (right) for the Charged Current interaction channel as a 
			function of final state charged lepton energy is shown for various choices incident neutrino 
			or 
			antineutrino energy. The difference between neutrino and antineutrino interaction strengths 
			(arising from the sign of $W_3$ term in Eq.\eqref{eq:diff}) is apparent.  At these energies, 
			only 
			the electron flavor participates in these 
			interactions. One can infer that final state lepton carries a significant portion of the incident 
			energy.}
	\end{figure}
	
	\subsection{Supernova neutrino fluence and oscillation physics}
	A supernova emits about 99\% of its gravitational binding energy into neutrinos over about 10 
	seconds~\cite{Colgate:1966ax}. Predicting the neutrino output of a wide range of supernovae 
	remains an outstanding problem in astrophysics~\cite{Janka:2016fox}. To obtain an estimate of 
	the total events in our detector,  we use a provisional model of the supernova neutrino fluence, 
	where the spectral number-fluence for flavor $\alpha$ is~\cite{Keil:2002in}  
	\begin{equation}\label{eq:fluence}
		F_\alpha^{(0)} = \frac{2.35 \times 10^{13}}{\text{cm}^2\,\text{MeV}} \cdot \left(\frac{{\cal 
		E}_\alpha}{10^{52}\,{\rm erg}}\right)\cdot\left(\frac{10\,{\rm kpc}}{d}\right)^2 \cdot 
		\frac{E_\alpha^3}{ \langle E_\alpha \rangle^5} \cdot  \exp \left(  -\frac{4 E_\alpha}{\langle 
		E_\alpha \rangle}\right)\,.
	\end{equation} 
	Here, $\mathcal{E_\alpha}$ is the net energy emitted per flavor, $d$ is the distance to the 
	supernova, and $E_\alpha$ (and $\langle E_\alpha\rangle$) are neutrino energies (and their 
	averages) in MeV \cite{Dasgupta:2011wg}. Note that the spectral number-fluence $F$ 
	gives the 
	number of neutrinos reaching the detector per unit surface area per unit 
	neutrino energy, \emph{without} including effects of flavor conversion. \\
	
	\begin{table}[b]
		\caption{\label{tab:Ealpha} Numerical values of $\langle E_\alpha \rangle$ in MeV for the 
			neutrino flavors for two benchmark models of supernova neutrino fluence. The 
			model dubbed High is an optimistic choice with larger event rate, compared to Low which is 
			a 
			conservative choice for $\langle E_\alpha \rangle$.}
		\centering
		\begin{tabular}{c c c c }
			\toprule
			Fluence Model & $\langle E_{\nu_e} \rangle$ &  $\langle E_{\bar{\nu}_e} \rangle$& $\langle 
			E_{\nu_x, 
				\bar{\nu}_x}  \rangle$ \\
			%	$\langle E_{\bar{\nu}_x} \rangle$ \\
			\midrule
			High (optimistic) & 12  & 15  & 18 \\
			Low (conservative)& 10 & 12 & 15 \\
			\bottomrule
		\end{tabular}
	\end{table}
	
	For our numerical estimates we assume equipartition of energies between all flavors, with 
	$\mathcal{E}_\alpha = 5\times 10^{52}$\,erg for all flavors. The distance $d$ is taken as 
	$10$\,kpc. The values of $\langle E_\alpha \rangle$ are model dependent and we take two sets 
	of 
	benchmark values for our analysis as outlined in Table \ref{tab:Ealpha}. The optimistic model, 
	with 
	larger values of $\langle E_\alpha\rangle$, gives a higher event rate and is dubbed as High in this 
	paper. A relatively conservative model, with smaller values of $\langle E_\alpha\rangle$, gives a 
	lower event rate and is dubbed as Low. \\
	
	We now include the effects of neutrino oscillations and flavor conversions. We assume that the 
	$\mu$ and $\tau$ flavors inside a supernova are identical, and thus the relevant conversion is 
	between the $e$-flux and that of some combination of $\mu$ and $\tau$, which we denote as 
	$x$~\cite{Akhmedov:2002zj}. The flavor-converted $\nu_e$ fluence at Earth can be 
	written as
	\begin{equation}
		F_e = p\,F_e^{(0)}+(1-p)F_{x}^{(0)}\,,
	\end{equation}
	where $p$ is the net survival probability of $\nu_e$ from its production to detection. For 
	$\bar{\nu}_e$, one has the analogous expression with $\bar{p}$,
	\begin{equation}
		F_{\bar{e}} = \bar{p}\,F_{\bar{e}}^{(0)}+(1-\bar{p})F_{\bar{x}}^{(0)}\,.
	\end{equation}
	For the NC events these oscillations make no difference. However, the CC events are sensitive 
	to 
	$p$ and $\bar{p}$. \\
	
	We will show our results for two oscillation scenarios: \emph{no oscillation}, where the fluences 
	remain unaltered and one has $p=1$ and $\bar{p}=1$, and \emph{flavor equilibrium}, where 
	$F_{\alpha} = \frac13 \left( F_{e}^{(0)} +  F_{\mu}^{(0)} +  F_{\tau}^{(0)} \right)$ so that 
	$p=\frac13$ and 
	$\bar{p}=\frac13$. The former is not physically realizable because MSW effects are guaranteed 
	\cite{Dighe:1999bi}. However, collective 
	oscillations~\cite{Pantaleone:1992eq,Pastor:2001iu,Duan:2005cp,Sawyer:2015dsa} inside the 
	star lead to intricate redistribution of flavor across all 
	energies~\cite{Duan:2006an,Fogli:2007bk,Dasgupta:2009mg}, perhaps even to almost 
	complete depolarization (i.e., equilibration between flavors) limited by conservation of lepton 
	numbers~\cite{Bhattacharyya:2020jpj, Johns:2019izj}. We sidestep the complexities by 
	restricting ourselves to a parametric study. For any other oscillation 
	scenario defined by the respective energy-dependent values of $p$ and $\bar{p}$, one can 
	obtain the predictions by 
	linearly combining the results of these two scenarios as
	\begin{equation}
		F_e = \frac{3}{2}(1-p)F^{{\rm fl.\,eqbm.}}_e + \frac{1}{2}(3p-1)F^{{\rm no\,osc.}}_e\,
		\label{oscp}
	\end{equation}
	and
	\begin{equation}
		F_{\bar{e}} = \frac{3}{2}(1-\bar{p})F^{{\rm  fl.\,eqbm.}}_{\bar{e}} + \frac{1}{2}(3\bar{p}-1)F^{{\rm 
				no\,osc.}}_{\bar{e}}\,.
		\label{oscpb}
	\end{equation}

	\section{Total number of events} \label{tot}
	
	For NC interaction, i.e., $ \overset{\scriptscriptstyle(-)}{\nu} d \rightarrow  
	\overset{\scriptscriptstyle(-)}{\nu}np$, the final state neutrino carries a large fraction of the 
	incident neutrino energy but cannot be detected as it does not further interact inside the 
	detector. The final state neutron and proton have kinetic energy that is approximately
	\begin{equation}
		T_{n(p)}  \sim \frac{E_\nu^2}{M_{n(p)}} \approx 240~\text{keV}~\left(\frac{E_\nu}{15\text{ 
		MeV}} \right)^2.
	\end{equation}
	The neutron can be captured on a nucleus, but its energy cannot be reconstructed. However, it 
	allows one to count the total number of NC events which is given by 
	\begin{equation}
		\mathcal{N}_{\text{NC}}  = N_d~\Delta t~\epsilon_n \int_{0}^{\infty} dE_\nu^\prime~ 
		\int_{0}^{\infty} dE_\nu~ \frac{d\phi}{d E_\nu}~\frac{d\sigma_{\text{NC}}}{dE_\nu^\prime}\,,
	\end{equation}
	where $N_d$ is the number of deuterons in the detector, $\epsilon_n$ is the efficiency of 
	neutron capture, and $\phi$ is the time-averaged neutrino flux with the dimensions of 
	cm$^{-2}$s$^{-1}$ over the duration of supernova neutrino emission, $\Delta t \approx 10$\,s. 
	In terms of the fluence ($F \equiv \Delta t \times d\phi/dE_\nu $), we can write 
	\begin{equation}
		\label{eq:nc}
		\mathcal{N}_{\NC}  = N_d~ \int_{0}^{\infty} dE_\nu~ F~\sigma_{\NC}\,,
	\end{equation}
	which can be evaluated using the tabulated values of the cross section by NGSK 
	\cite{Nakamura:2000vp}. We assume that the neutron capture efficiency is $\epsilon_n=1$. We 
	discuss this assumption in more detail in Section\,\ref{conc}. The quenched scintillation signal 
	from the proton can be detected and carries spectral information, but only a fraction of these 
	events will be above the threshold. This is studied in detail in Section\,\ref{protrec}.  \\

	For CC interactions, we can not only reconstruct the electron and positron energy, but also the 
	interaction point inside the detector. The binned event rate can be used to obtain the shape of 
	the incoming neutrino spectrum. The number of events in a bin $j$ is, 
	\begin{equation}
		\mathcal{N}_j  = N_d ~\Delta t \int_{\Delta E_e^\prime}^{} dE_e^\prime~\varepsilon_e ~\left[ 
		\int_{0}^{\infty} dE_\nu~ \frac{d\phi}{d E_\nu}~\frac{d\sigma_{\CC}}{d E_e^\prime} \right]\,,
	\end{equation}
	where $\varepsilon_e$ is the efficiency of detecting the electrons. For the remainder of the 
	discussion, we have not included the detector effects and assume $\varepsilon_e = 1$. \\
	
	Instead of binned events, one can ask the total number of CC events that will be observed by 
	such a detector. Even though $E_e^\prime$ depends on $E_\nu$, one can carefully use the 
	expression 
	\begin{equation}
		\mathcal{N}_{\CC}  = N_d~\Delta t~\int_{0}^{\infty} dE_e^\prime ~ \int_{0}^{\infty} dE_\nu~ 
		\frac{d\phi}{d E_\nu}~\frac{d\sigma_{\CC} }{d E_e^\prime} \equiv \sum_{j}^{}\mathcal{N}_{j}\,,
	\end{equation}
	to calculate the total number of events. After  accounting for the kinematic limits, one can write 
	$\mathcal{N}_{\CC} $ in terms of fluence as,
	\begin{equation}
		\mathcal{N}_{\CC}  = N_d~ \int_{0}^{\infty} dE_\nu~ F~\sigma_{\CC}\,, 
	\end{equation}
	which is similar to the one obtained for NC interaction \eqref{eq:nc}. One can use the tabulated 
	cross-sections in NSGK \cite{Nakamura:2000vp} to estimate the event rates.\\
	
	Important interactions of neutrino with the deuteron in a DLS are mentioned in the Table 
	\ref{tab:interactions}. The number of target particles ($N_d$) is given for 1 kton of detector 
	volume (see Appendix\,\ref{estNd} for estimates). Although the flux of neutrinos from supernova 
	typically peaks around 10 MeV, the cross section approximately scales as $E_\nu^{2.5}$ and 
	most of the events arise from $E_\nu \sim 20$ MeV. We have mentioned the cross section for 
	the interaction channel at 20 MeV to understand their relative strengths.\\
	
	\begin{table}[!b]
		\caption{\label{tab:interactions} Representative cross sections for the four neutrino induced 
		deuteron dissociation channels along with their Q values.}
		\centering
		\begin{tabular}{  c   c   c   c     }
			\toprule
			Interaction & Channel & -Q (MeV) & $\sigma (E_\nu = 20~\text{MeV})$ \\
			\midrule
			$\nu + d \rightarrow \nu + n + p$ & NC & 2.224 &$6.98 \times10^{-42}$ cm$^2$   \\
			$\overline{\nu} + d \rightarrow \overline{\nu} + n + p$ & NC& 2.224 & $6.28 
			\times10^{-42}$ 
			cm$^2$  \\
			$\nu_e + d \rightarrow e^- + p + p$ & CC & 1.442& $15.61 \times10^{-42}$ cm$^2$   \\
			$\overline{\nu}_e + d \rightarrow e^+ + n + n$ & CC & 4.028& $9.54 \times10^{-42}$ 
			cm$^2$   \\ 
			\bottomrule
		\end{tabular}
	\end{table}

	\begin{table}[t]
		\caption{\label{tab:NCcount} Number of NC events for each flavor in 1 kton of DLS from a 
		fiducial 
			supernova at 10 kpc. The event numbers for individual flavors are not observable (as they 
			cannot 
			be 
			distinguished) but provided to allow cross checks, and one only measures the total. The 
			subscript 
			$x$ refers to any of $\mu,\tau$; note that the $\bar{\nu}_x$ event rates are 
			slightly lower owing to their smaller NC cross section. The events are reported for different 
			scenarios depending on high / low fluence model and no oscillation / flavor equilibrium case.}
		\centering
		\begin{tabular}{ c   c   c  c   c  c }
			\toprule 
			Scenario & $\nu_e$ & $\bar{\nu}_e $& $\nu_x$ & $\bar{\nu}_x$ & Total (rounded off)\\
			\midrule
			High; no oscillation & 48.4 & 59.7 & 88.1 & 75.6 & 435 \\
			High; flavor equilibrium &74.8 & 70.3 & 74.8 & 70.3 & 435\\
			Low; no oscillation & 36.1 & 43.7 & 67.8 & 59.7 & 335 \\
			Low; flavor equilibrium & 57.3 & 54.3 & 57.3 & 54.3 & 335\\
			\bottomrule
		\end{tabular}
		
	\end{table}
	
	\begin{table}[h]
		\caption{\label{tab:CCcount} Number of CC events for $\nu_e$ and $\bar{\nu}_e$ in 1 kton of 
			DLS from a fiducial supernova at 10 kpc. The two flavors can, in principle, be distinguished 
			in the 
			detector by tagging the hadrons in the final state. The events are reported for different 
			scenarios depending on high / low fluence model and no oscillation / flavor equilibrium 
			case.} 
		\centering
		\begin{tabular}{ c   c   c }
			\toprule
			Scenario & $\nu_e$ & $\bar{\nu}_e $\\
			\midrule
			High;  no oscillation & 111 & 90 \\
			High; flavor equilibrium & 170 & 108 \\
			Low; no oscillation & 84 & 63 \\
			Low; flavor equilibrium & 130 & 81 \\
			\bottomrule
		\end{tabular}
	\end{table}
	
	We have reported the forecasted NC and CC event rates for the ``no oscillation'' and ``flavor 
	equilibrium'' scenarios in Tables \ref{tab:NCcount} and \ref{tab:CCcount}. For NC interaction, we 
	see that the total number of events remains unchanged and are independent of the oscillation 
	scenario. For CC interaction, the events are larger as we move away from the no oscillation 
	scenario because of the following reason. The $\nu_{\mu, \tau}$ are emitted at a slightly larger 
	temperature and a fraction of these are converted to $\nu_e$ due to oscillations. As the 
	interaction cross section increases with neutrino energy, a significant number of events is 
	obtained 
	from these neutrinos. Similar event rates for SNO were estimated in Ref. \cite{Dutta:2001nf}.\\
	
	As noted before, one can easily use these event rates to get the results for any other oscillation 
	scenario. For example, with only adiabatic MSW effects in the inverted (or normal) mass ordering 
	\cite{Dighe:1999bi}, one has $p = \sin^2\theta_{13} \approx 0.02$ (or 
	$\cos^2\theta_{13}\sin^2\theta_{12} \approx 0.3$) and 
	$\bar{p}=\cos^2\theta_{13}\cos^2\theta_{12} \approx 0.68$ (or $\sin^2\theta_{13} \approx 
	0.02$)~\cite{Zyla:2020zbs}. The relevant event rates can be found by combining the  ``no 
	oscillation'' and ``flavor equilibrium'' event rates using Eqs.\,\eqref{oscp} and \eqref{oscpb}. For 
	example, with fluence model High, one finds 197 (172) CC events for $\nu_e$ and 99 (81) events 
	for $\bar\nu_e$.\\

	\section{Energy spectrum of events} \label{spec}
	
	In the previous section, we have seen that the 1 kton of DLS can see several hundreds of 
	neutrino events from a supernova in the galaxy. We have also mentioned that the interactions of 
	the neutrino on deuteron have unique signatures and one can distinguish the events. In this 
	section, we look at the spectrum of events for each channel. 
	
	\subsection{Neutral current interactions}\label{protrec}
	
	The NC mediated neutrino dissociation of the deuteron is given by the reaction
	\begin{equation}
		\overset{\scriptscriptstyle(-)}{\nu} + d \rightarrow \overset{\scriptscriptstyle(-)}{\nu} + n + p\,,
	\end{equation}
	where $\overset{\scriptscriptstyle(-)}{\nu}$ includes all three flavors of (anti)-neutrino. The 
	neutron 
	in the final state leads to observable photons via neutron capture, and we used it to estimate the 
	total event rate. Now we focus on the proton.\\
	
	\subsubsection{Quenched proton spectrum}
	
	The final state proton loses its kinetic energy mainly due to ionization energy losses which is 
	converted into scintillation photons in the detector (cross section for the alternate process, 
	capture of proton on deuteron emitting 5 MeV gamma rays, is small\,\cite{Schadow:1998hv} and 
	the event rate is negligible). The total amount of scintillation light is not the same as the total 
	kinetic energy due to photo-saturation of the detector, also known as \emph{quenching}. 
	Beacom, 
	Farr, and Vogel (BFV) \cite{Beacom:2002hs} proposed that the quenched proton signal from 
	neutrino 
	proton elastic scattering can be used to detect supernova neutrinos in scintillation detector such 
	as KamLAND and Borexino. A method for reconstructing the supernova spectrum was proposed 
	in 
	\cite{Dasgupta:2011wg}. See also Refs.~\cite{Lu:2016ipr, Li:2017dbg, Li:2019qxi, 
	Nagakura:2020bbw}.\\
	
	In contrast to BFV, where there was only one hadron in the final state, here one has an 
	additional particle, a neutron, in the final state. If the scintillation from the proton can be tagged 
	using the $\gamma$ photons from the neutron capture, one can achieve significant reduction in 
	backgrounds. Moreover, the kinematics of the outgoing proton is not identical to that in BFV as 
	this channel has a three-body final state. However, a simplification is possible if the deuteron 
	breakup occurs in two stages - $\nu d \rightarrow \nu (d^\ast \rightarrow np)$ where $d^\ast$ is 
	an unstable excited state of the deuteron. In this picture, one can neglect the difference 
	between proton and the neutron, as $T_p - T_n \ll M_d $. We have checked that the differential 
	cross section obtained by this method agrees with the plots presented by NSGK.\\
	
	The differential cross section with respect to the proton recoil energy ($T_p$) can be obtained 
	using
	\begin{equation}
		\label{diff}
		\frac{d \sigma}{dT_p} = \left|  \frac{\partial E_\nu^\prime}{\partial T_p} \right|  
		\frac{d\sigma}{dE_\nu^\prime}\,,
	\end{equation}
	where we use the energy conservation relation $E_\nu^\prime = E_\nu - B - 2 T_p$ and $B=$ 
	2.223 MeV is the binding energy of the deuteron. The numerical values of the differential cross 
	section for the relevant range of values of $E_\nu$ are tabulated in Appendix\,\ref{tabxsec}. \\
	
	\begin{figure}[t]
		\centering
		\includegraphics[width=7.19cm]{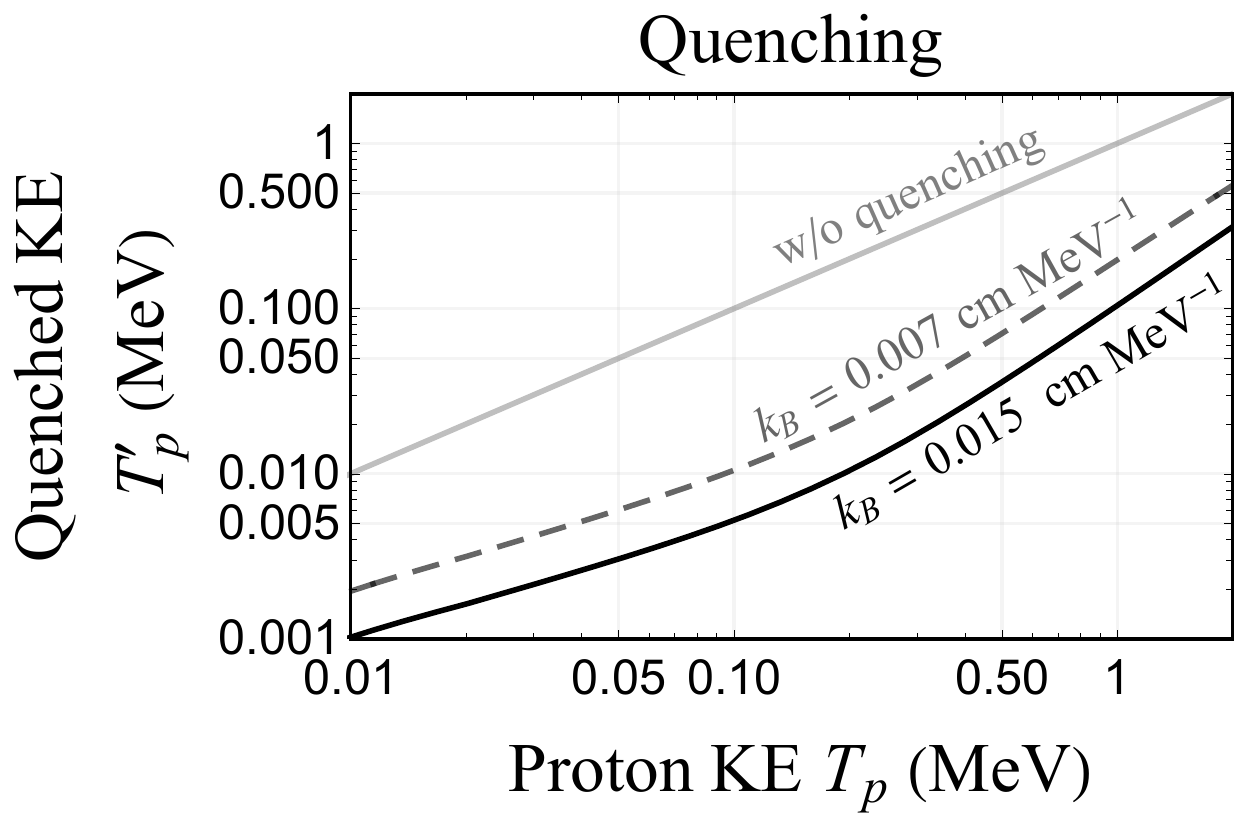}
		\includegraphics[width=7cm]{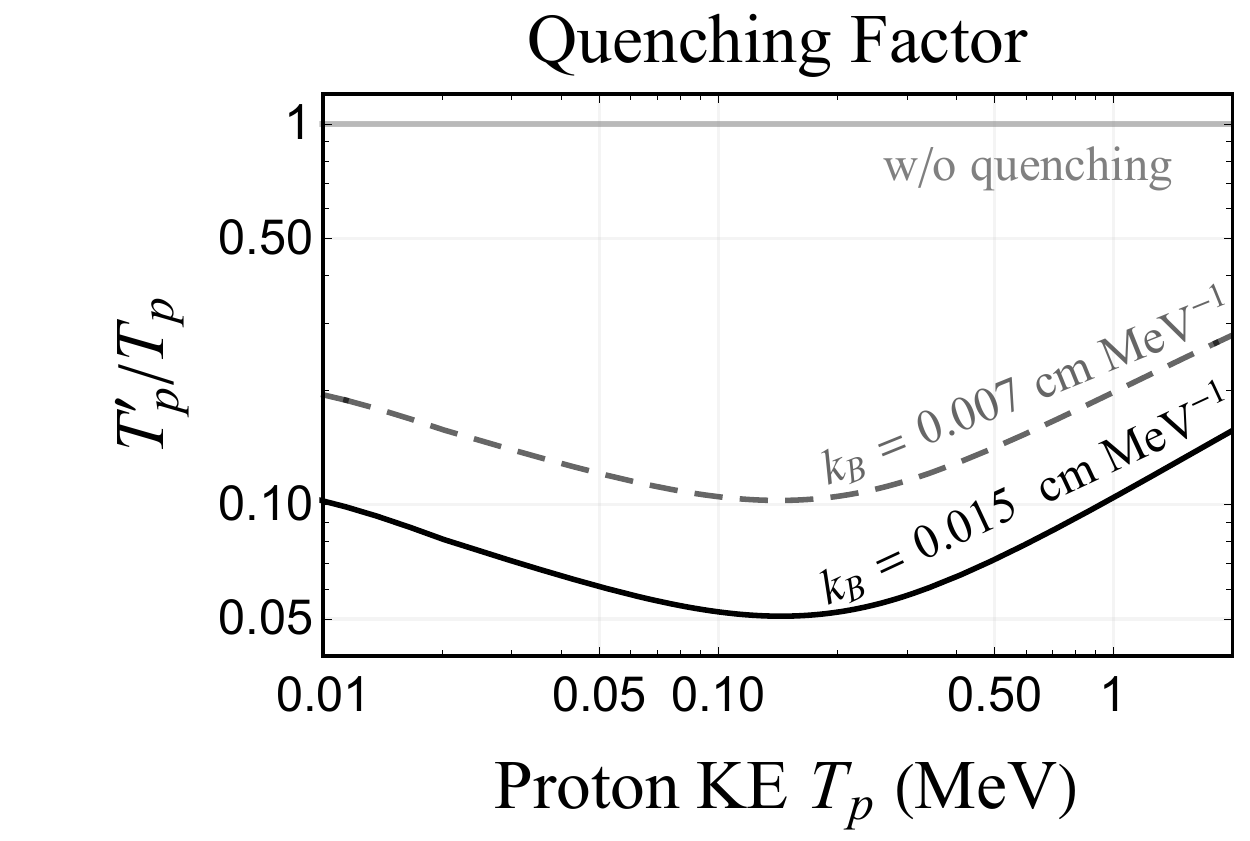}
		\caption{\label{fig:quenching} The quenched kinetic energy (left) and the quenching factor 
			(right) is shown as a function of the proton energy for two benchmark values of the Birks 
			constant. The solid curve is for $k_B = 0.015$\,cm\,MeV$^{-1}$ and the dashed curve for 
			$k_B 
			= 0.007$\,cm\,MeV$^{-1}$. We also show the unquenched scenario $T_p^\prime = T_p$.}
	\end{figure}

	A proton with recoil energy $T_p$ will lose energy in the detector due to collisions with electrons 
	and the nucleus. It is seen that in typical materials the energy loss rate is dominated by 
	scattering 
	with electrons and approximately is $(60 - 300)$ MeV/cm for $T_p \approx$ few MeV. Accurate 
	numerical values for the energy loss rate can be obtained from the PSTAR 
	database\footnote{\url{www.physics.nist.gov/PhysRefData/Star/Text/PSTAR.html}} for various 
	targets. 
	Similar data can also be obtained using SRIM\footnote{\url{www.srim.org}} 
	\cite{ziegler2008srim}. 
	Only 
	a small part of the recoil energy, as compared to that of an electron of the same energy, is 
	converted into scintillation light. This electron equivalent quenched kinetic energy $T_p^\prime$ 
	is 
	given by 
	\begin{equation}
		T_p^\prime = Q(T_p) = \int_{0}^{T_p} \frac{dT}{1 + k_B  \langle dT/dx \rangle}\,,
	\end{equation}
	where $k_B \approx 0.015$ cm\,MeV$^{-1}$ is the Birks constant \cite{Birks_1951} and $\langle 
	dT/dx \rangle $ is the average stopping power that depends on $T_p$. For an accurate 
	estimate, 
	one must take the weighted average energy loss due to carbon and deuteron. Unfortunately, the 
	stopping power of deuteron is not known and we approximate it by the stopping power of 
	hydrogen. The stopping power of  $\text{C}_n \text{D}_{2n} $ is thus taken to be
	\begin{equation}
		\bigg \langle \frac{dT}{dx} \bigg \rangle = \sum_{i} w_i\bigg \langle \frac{dT}{dx} \bigg \rangle_i 
		\approx 0.749\bigg \langle \frac{dT}{dx} \bigg \rangle_\text{C} + 0.251\bigg \langle 
		\frac{dT}{dx} \bigg \rangle_\text{H}.
	\end{equation}
	The Birks constant for DLS is also unknown and hence taken to be of same order as other 
	scintillator materials. The quenching factor also depends on the density of the medium which is 
	also an unknown. For simplicity, we assume it to be 1 g/cm$^3$ and note that the change due 
	to 
	density can be effectively captured by changing the Birks constant. In Figure \ref{fig:quenching} 
	we 
	have shown the quenched kinetic energy of the proton for two choices $k_B = 
	0.007$\,cm\,MeV$^{-1}$ and $k_B = 0.015$\,cm\,MeV$^{-1}$. To obtain conservative 
	estimates 
	of the event rates, we have used the latter for remainder of this paper.\\
	
\begin{figure}[t]
		\centering
		\includegraphics[width=7cm]{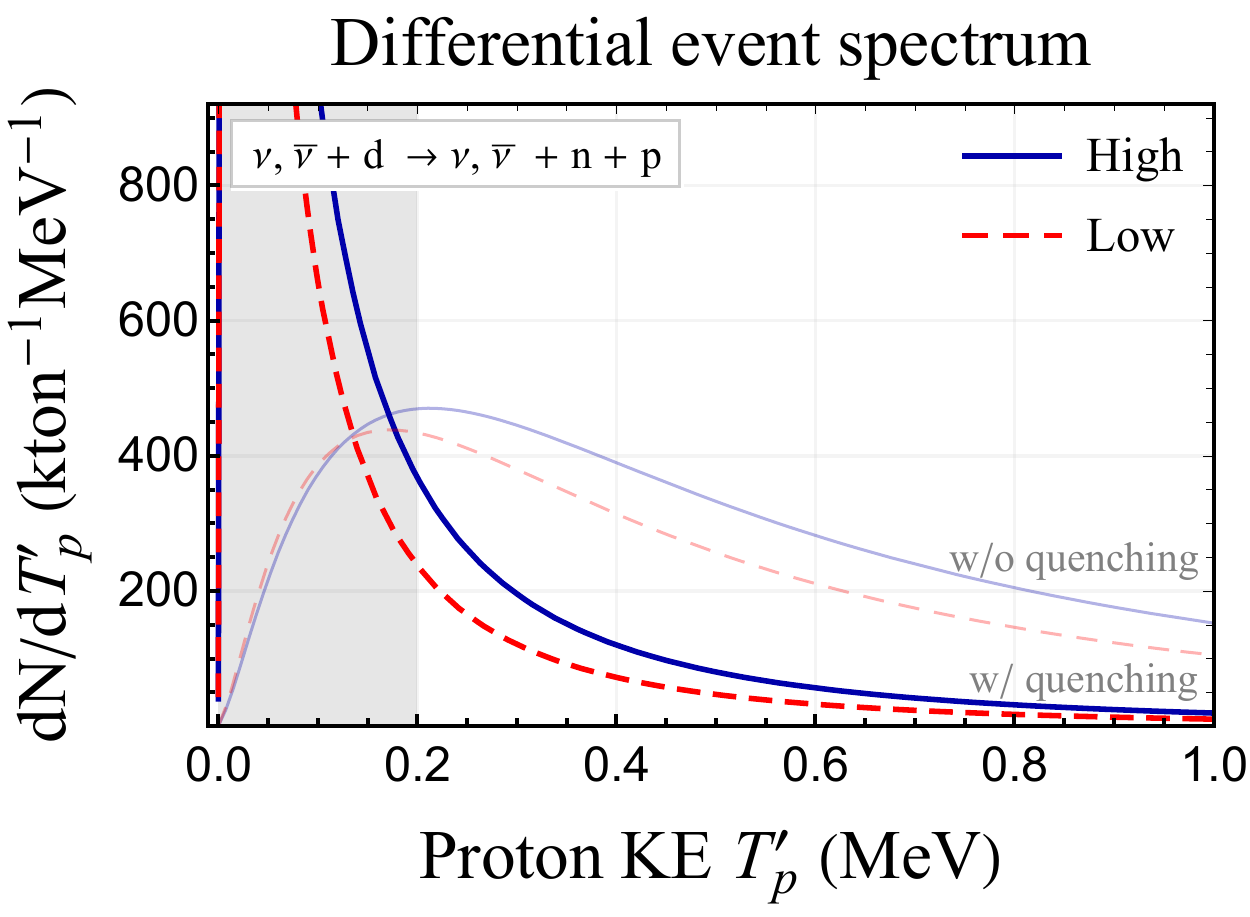}\hspace{.5cm}
		\includegraphics[width=6.9cm]{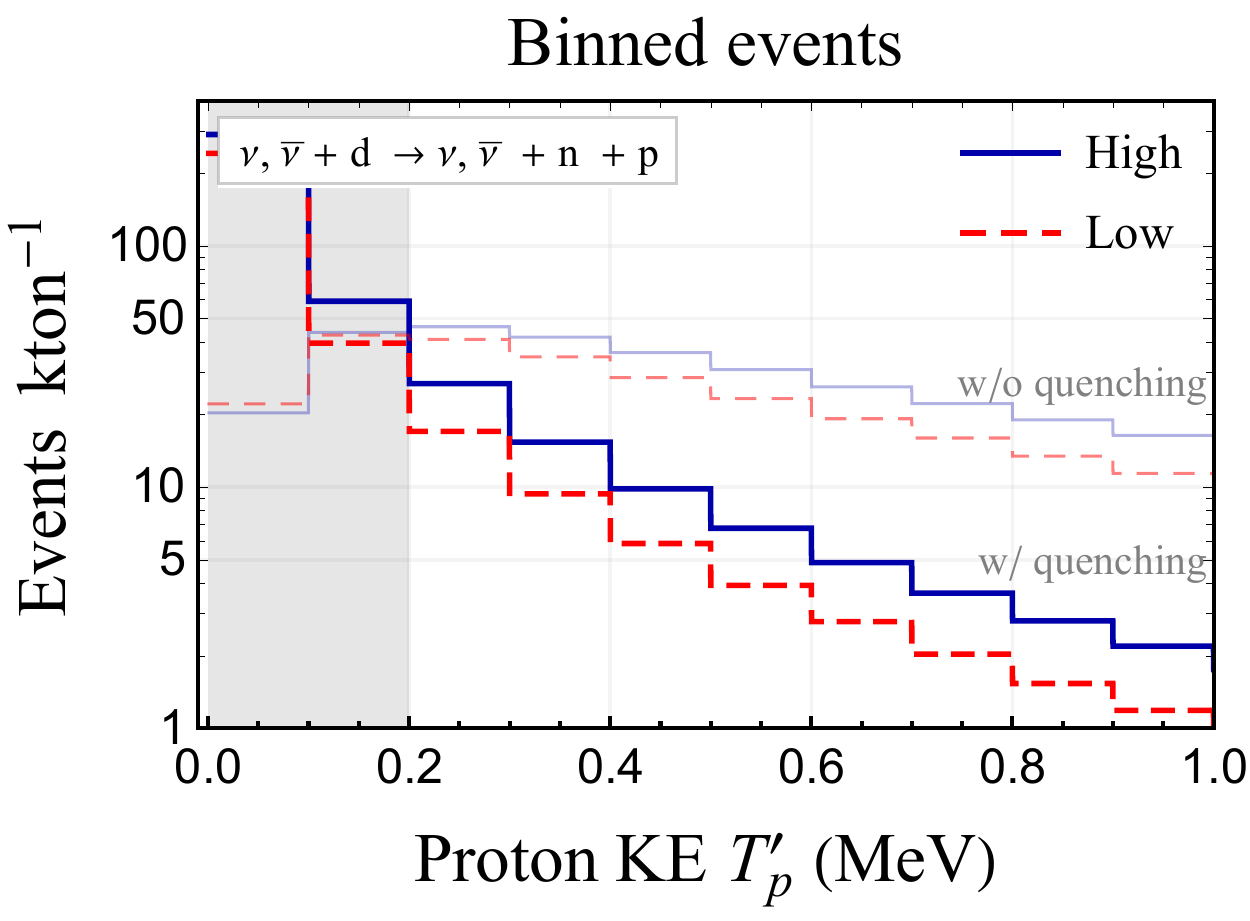}
		\caption{\label{fig:quenched}  The differential event spectrum (left) and binned events (right) 
			with proton recoil energy with and without quenching are shown. The solid blue curves are 
			for 
			optimistic fluence model High and dashed red curves are for the conservative fluence 
			model 
			Low. The gray shading shows energies below threshold, where we take $E_{th}$ = 200 keV 
			due 
			to $^{14}$C beta decay background. However, this threshold can be lowered by tagging 
			the 
			final state neutron.}
	\end{figure}

	We can now estimate the quenched proton event spectrum using, 
	\begin{equation}
		\label{eqn:quench}
		\frac{d \mathcal{N}}{dT_p^\prime}  = \frac{N_d}{dT_p^\prime/dT_p}~ \int_{0}^{\infty} dE_\nu~ 
		F~\frac{d\sigma}{dT_p}\,,
	\end{equation}
	where the differential cross section is estimated using Eq. \eqref{diff}. In Figure 
	\ref{fig:quenched} we have shown the differential event rate with proton recoil energy $T_p$ as 
	well as the quenched kinetic energy $T^\prime$. The binned event rate ($\Delta E_{\rm bin}$ = 
	0.1 MeV) with proton recoil energy $T_p$ and the quenched kinetic energy $T^\prime_p$ is also 
	shown. \\
	
	In ordinary scintillator detectors,  the dominant background at these energies comes from the 
	beta 
	decay of $\ce{^{14}_{}C}$. The low energy electrons (up to 200 keV) can mimic the proton 
	scintillation signal. The typical event rate for these backgrounds is $\mathcal{O}(10^7)$ events 
	per 
	day. Hence, we consider a threshold of 0.2 MeV similar to Ref.\,\cite{Dasgupta:2011wg} to get rid 
	of these backgrounds. One must note that the electron scintillation will not have an 
	accompanying 
	neutron capture signal. Hence these background events can, in principle, be distinguished from 
	the 
	signal events using appropriate tagging. This will lower the threshold and allow for better signal 
	to 
	noise ratio. For a DLS, dedicated study of backgrounds is needed.  In Table 
	\ref{tab:NCQuenched}, 
	we have shown the events for the proton recoil via NC channel for various thresholds for both 
	quenched and unquenched spectrum.
	
	\begin{table}[t]
		\caption{Total number of NC events per kton, without  quenching ($T_p^\prime = T_p$) and 
		with 
			quenching ($T_p^\prime = Q(T_p) $), for three choices of energy threshold $E_{th} =$ 10 
			keV, 100 
			keV, and 200 keV. The neutrino and antineutrino event rates are different, owing to their 
			different 
			cross sections in addition to the difference in fluxes of $\nu_e$ and $\bar\nu_e$, but only 
			the total 
			is observable. The energy spectra are shown in Figure \ref{fig:quenched}. The results 
			depend on 
			the fluence model but not on the oscillation scenario.}
		\centering
		\begin{tabular}{ c c c c}
			\toprule
			Fluence & Flavor & Events/kton without quenching& Events/kton  with quenching  \\
			model &  & $E_{th}=10\,\,;\,100\,\,;\,200\,({\rm keV})$& $E_{th}=10\,\,;\,100\,\,;\,200\,({\rm 
				keV})$  \\
			\midrule
			\multirow{3}{*}{High} & $\nu_i$& 74.7\,;\,71.3\,;\,64.0 &  64.6\,;\,25.0\,;\,14.8 \\
			& $\overline{\nu}_i$&69.4\,;\,66.3\,;\,59.4 & 60.0\,;\,22.7\,;\,13.3 \\
			&Total& 432.4\,;\,412.8\,;\,370.3 & 373.8\,;\,143.2\,;\,84.2   \\
			\midrule
			\multirow{3}{*}{Low} & $\nu_i$& 56.7\,;\,53.1,  46.1 & 46.6\,;\,15.4\,;\,8.6 \\
			& $\overline{\nu}_i$& 53.4\,;\,50.0\,;\,43.2 &  43.7\,;\,14.2\,;\,7.9 \\
			&Total& 330.6\,;\,309.3\,;\,268.0 & 271.2\,;\,89.0\,;\,49.4  \\
			\bottomrule
		\end{tabular}
		\label{tab:NCQuenched}
	\end{table}
	
	\subsection{Charged current interactions}
	As mentioned earlier, due to kinematic threshold only the electron flavor neutrino and 
	antineutrino participate via CC interaction for supernova neutrinos. The final state for $\nu_e$ is 
	different from $\overline{\nu}_e$ and the two interaction channels can be distinguished by 
	appropriate tagging. Hence, unlike the NC case, the two flavors are discussed separately as 
	follows. 
	
	\subsubsection{CC interaction of $\nu_e$}\label{CC_nu_e}
	The CC mediated $\nu_e$-deuteron interaction is given by
	\begin{equation}
		\nu_e + d \rightarrow e^- + p + p\,,
	\end{equation}
	where the final state electron is observed through its scintillation radiation. One 
	can, in principle, reconstruct the interaction point by analysing the lepton observed by 
	the PMTs and recover spectral information of the incident neutrino. Hence this channel is 
	important for reconstructing the supernova neutrino spectrum. The two recoiling protons will 
	generate a quenched scintillation signal as well. It seems, a priori, that the signals will overlap 
	and it 
	will be challenging to segregate and tag the events. A quantitative estimate of required detector 
	resolution calls for detailed numerical simulations that is beyond the scope of this paper and we 
	encourage further investigations. \\
	
\begin{figure}[t]
		\centering
		\includegraphics[width=7cm]{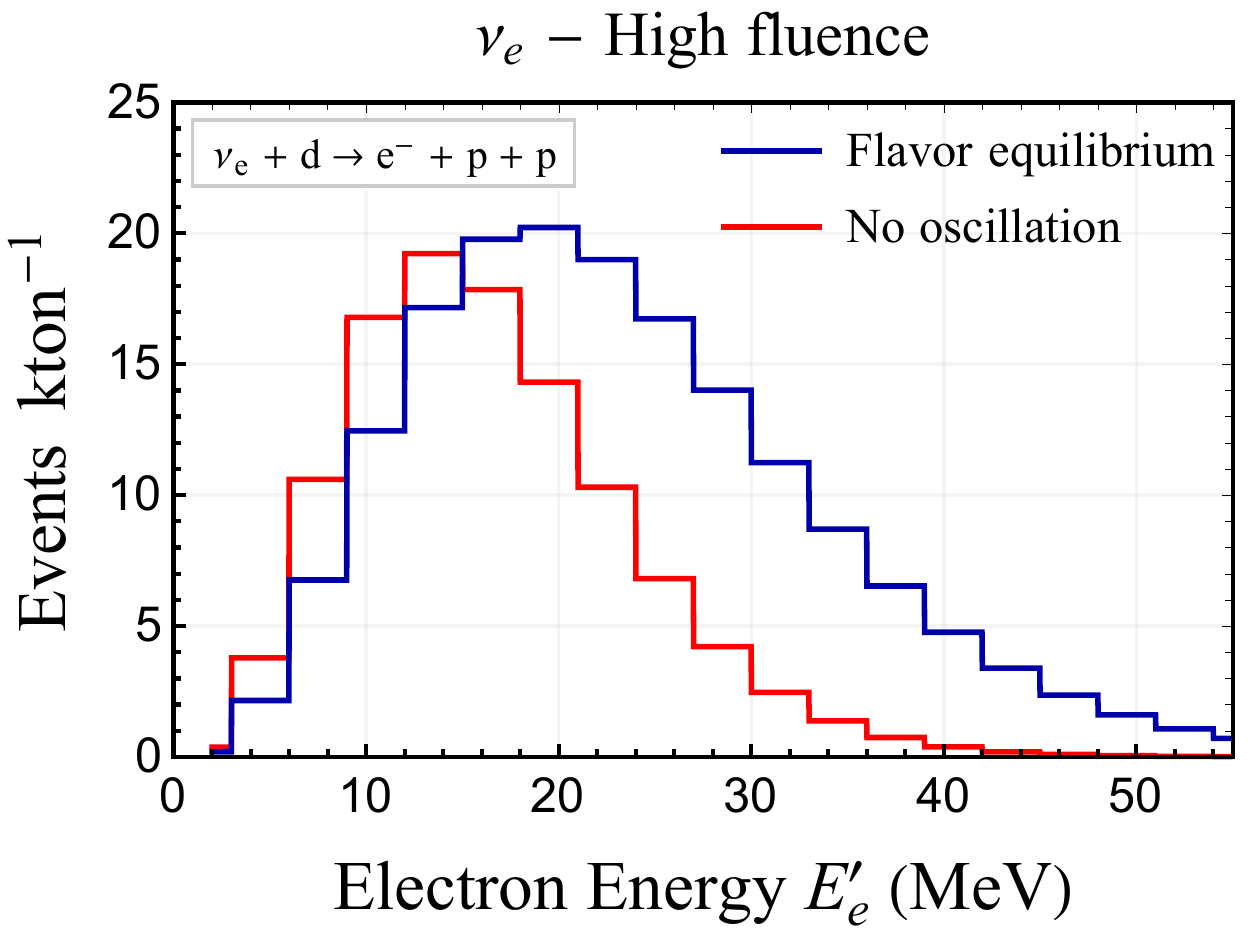}\hspace{0.5cm}
		\includegraphics[width=7cm]{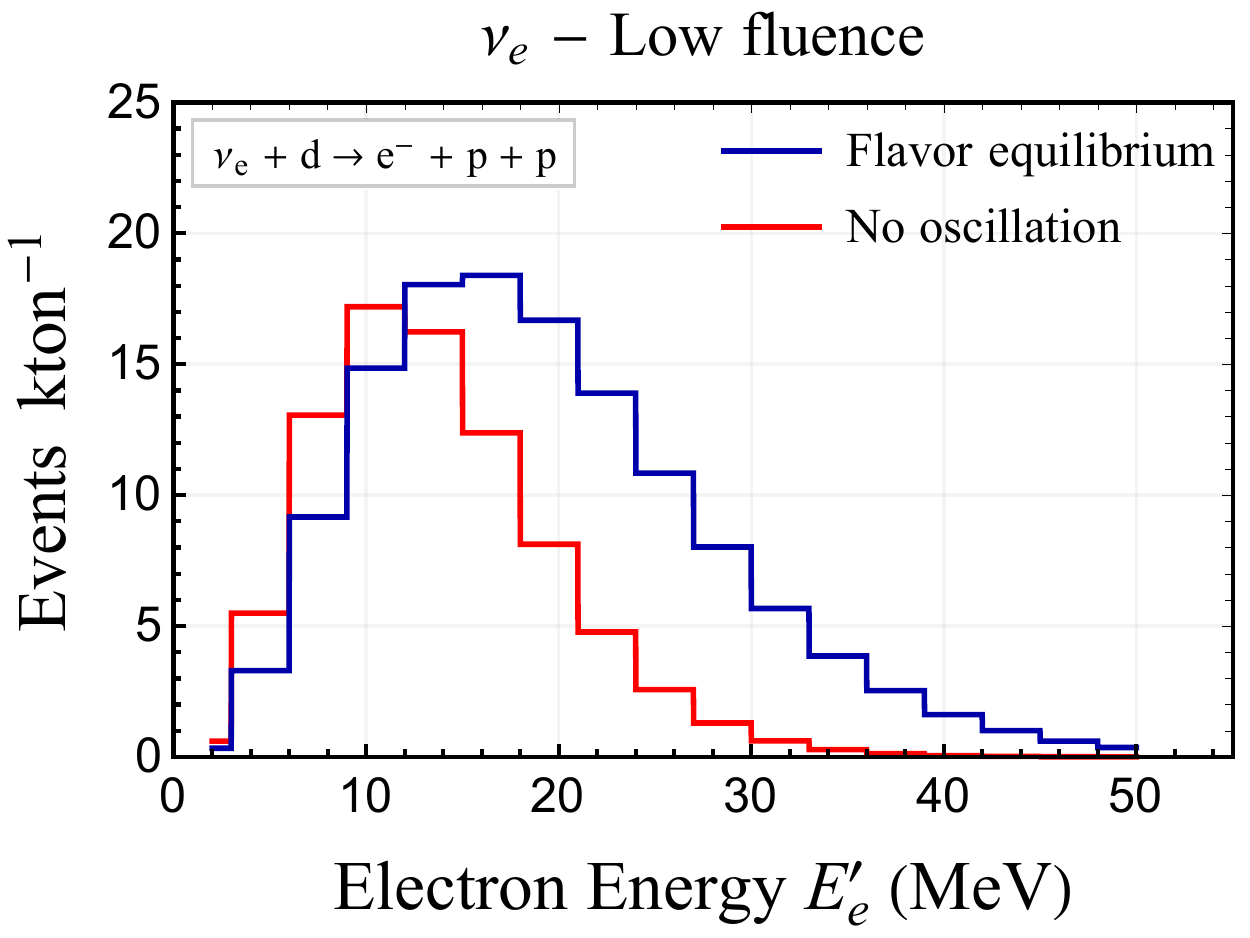}
		\caption{\label{fig:binned2} Binned event spectrum for $\nu_e$ detected via Charged Current 
		interactions, 
			with respect to the electron energy. The spectrum assuming flavor equilibrium 
			(no oscillation) is shown with blue (red) lines. Results are shown for the optimistic fluence 
			model High 
			(left) and conservative model Low (right).}
	\end{figure}

	We estimate the event rate binned with respect to observable electron energy. The bin size, 
	$\Delta E_{\rm bin} = 3$ MeV, is chosen slightly larger than the expected energy resolution so 
	that 
	the bins are uncorrelated. We show the results in Figure \ref{fig:binned2} for the two fluence 
	models considered in this paper. It must be noted here that DLS  can provide excellent resolution 
	of this channel, both in energy and time, that can only be achieved by a LAr detector such as 
	DUNE. 
	During the 15-20 ms of the neutronization burst, we expect 4.5 (1.5) events from this channel 
	for 
	the no oscillation (flavor equilibrium) scenario. The details are in Appendix 
	\ref{App:Burst}.
	\begin{figure}[t]
		\centering
		\includegraphics[width=7cm]{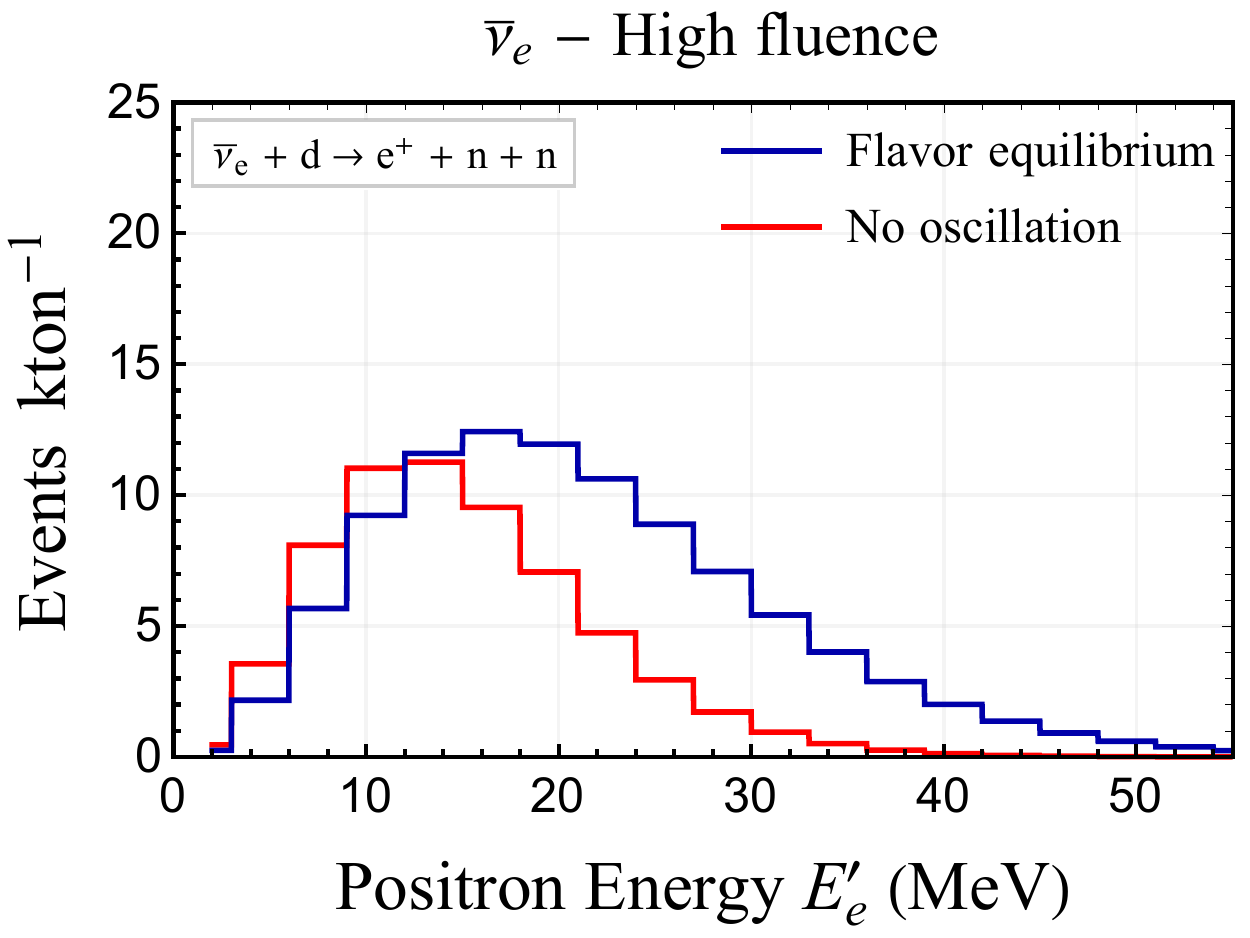}\hspace{0.5cm}
		\includegraphics[width=7cm]{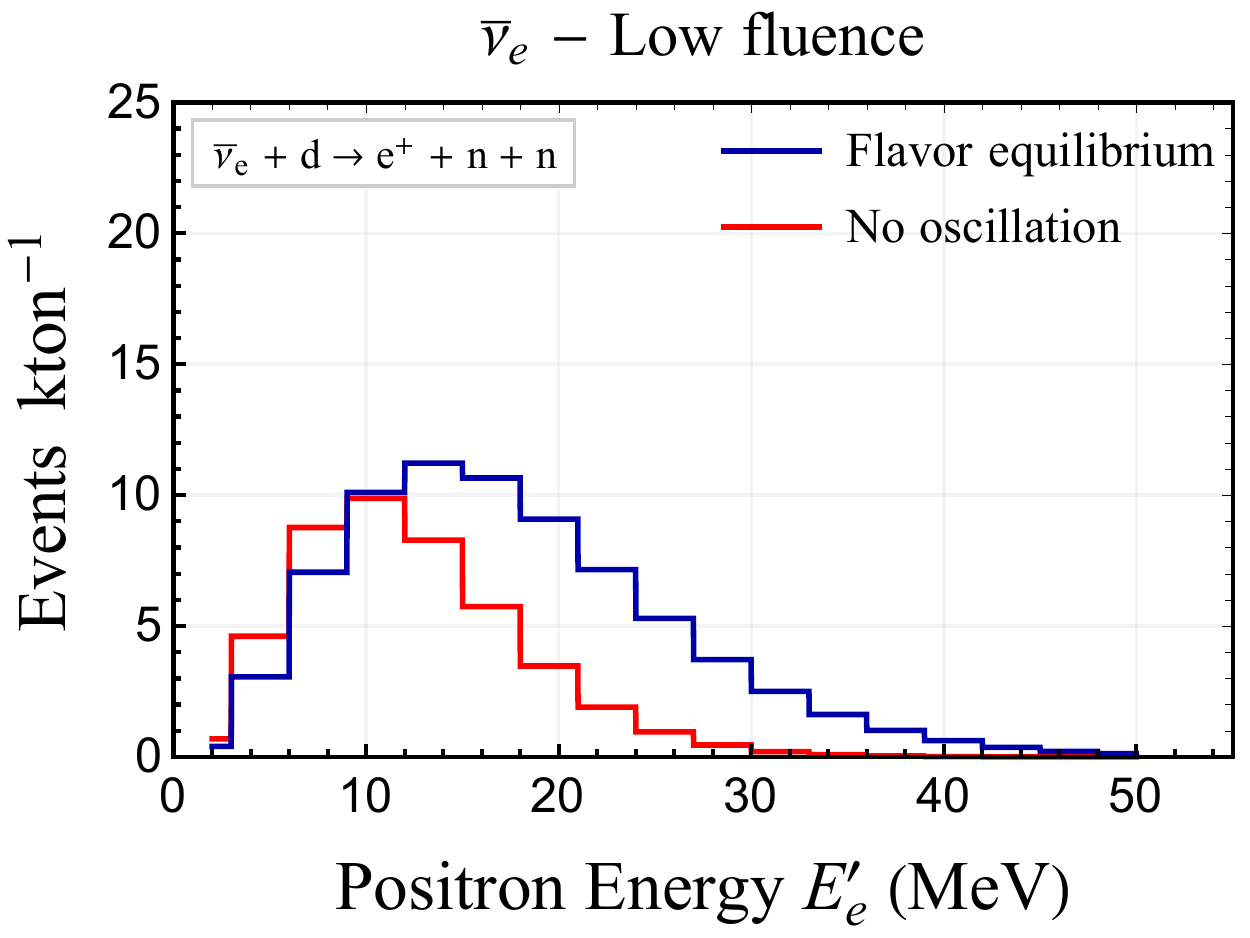}
		\caption{\label{fig:binned1} Binned event spectrum for $\bar\nu_e$ detected via Charged 
		Current interactions, 
			with respect to the positron energy. The spectrum assuming flavor equilibrium 
			(no oscillation) is shown with blue (red) lines. Results are shown for the optimistic fluence 
			model High 
			(left) and conservative model Low (right).}
	\end{figure}
	
	\subsubsection{CC interaction of $\overline{\nu}_e$}
	
	The CC mediated $\overline{\nu}_e$-deuteron interaction is given by
	\begin{equation}
		\bar{\nu}_e + d \rightarrow e^+ + n  + n\,,
	\end{equation}
	where the final state positron can be observed through its scintillation. The two neutrons in the 
	final 
	state will thermalize with the medium and their capture will give detectable photons. In principle, 
	the simultaneous observation of these neutrons will provide a cleaner tag for the event 
	compared 
	to the inverse beta decay reaction of $\bar{\nu}_e$ on a free proton. We have shown the binned 
	event spectrum for this interaction in Figure \ref{fig:binned1}. 
	
	\subsection{Elastic scattering off electrons}
	
\begin{figure}[h!]
		\centering
		\includegraphics[width=7cm]{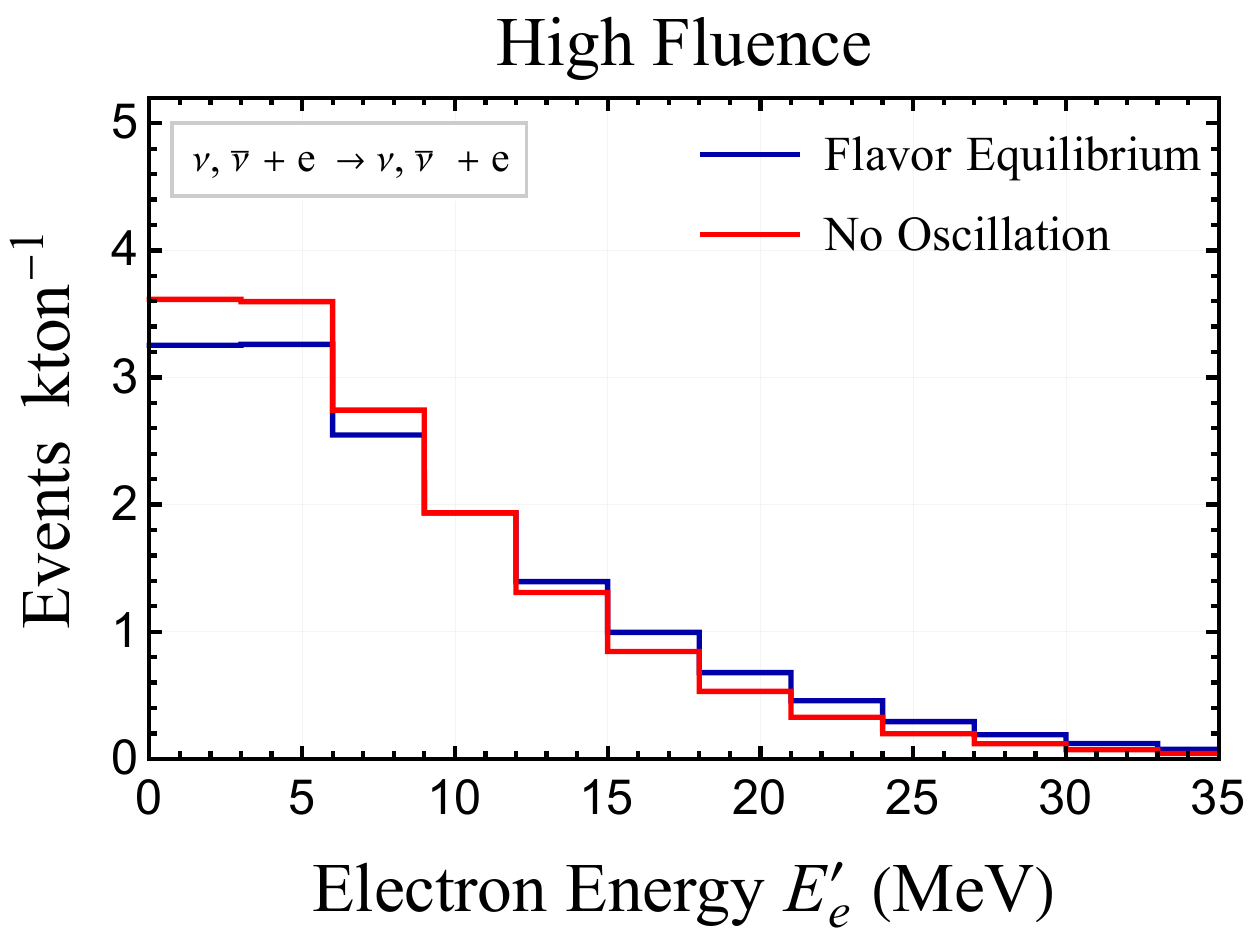}
		\includegraphics[width=7cm]{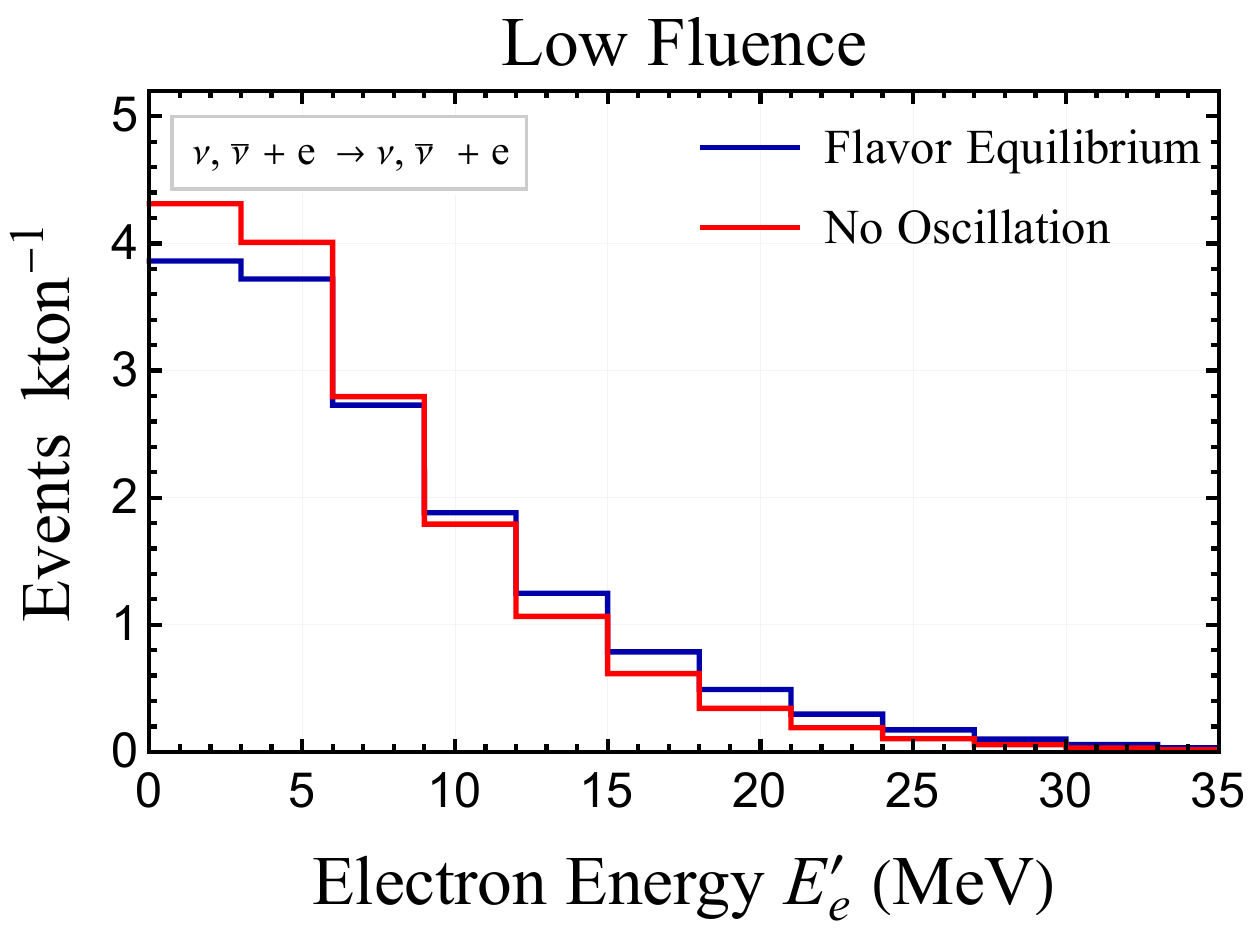}	
		\caption{\label{fig:ES} Binned event spectrum for elastic scattering off electrons with respect 
		to 
			the measured electron energy. Results are shown for the optimistic fluence model High 
			(blue) and conservative model Low (red). Both models results in an aggregate of 
			approximately 15 events.} 
	\end{figure}

	The elastic scattering (ES) of neutrinos off electrons in the detector is an important channel to 
	measure 
	NC events in detectors that mostly rely on IBD. In DLS, we expect an aggregate of 15 events, 
	mostly in 
	the lower energy bins due to zero threshold of the interaction. Unlike the NC interaction with 
	deuteron, the ES channel is sensitive to flavor conversion as the cross sections for electron and 
	non-electron flavors are different \cite{Marciano:2003eq}. In figure \ref{fig:ES}, we have shown 
	the binned spectrum of 
	events for the two fluence models considered in this paper. The scintillation from recoiling 
	electrons could be considered as a \emph{background} for reconstruction of the incident 
	neutrino 
	spectrum from the CC channel discussed in \ref{CC_nu_e}. 
	
	\section{Secondary interactions}\label{sec:Secondary}
	So far, we have focused on neutrinos scattering with deuteron which leads to protons and 
	neutrons in the final state. We have assumed that the protons lead to detectable scintillation, 
	and all of the neutrons are captured and detected. Due to their large energy loss rate, mainly 
	due to ionization, the protons travel $\mathcal{O}(0.1)$ mm in the detector. On the other hand, 
	the neutrons undergo multiple scatterings off the neighboring nuclei before getting captured. In 
	this section, we analytically examine the interactions of these recoiling neutrons in the 
	detector\footnote{We thank the anonymous referee for suggesting this.}. 
	
	\subsection{Fate of the recoiling neutron}
	
\begin{figure}[h!]
		\centering
		\includegraphics[width=9cm]{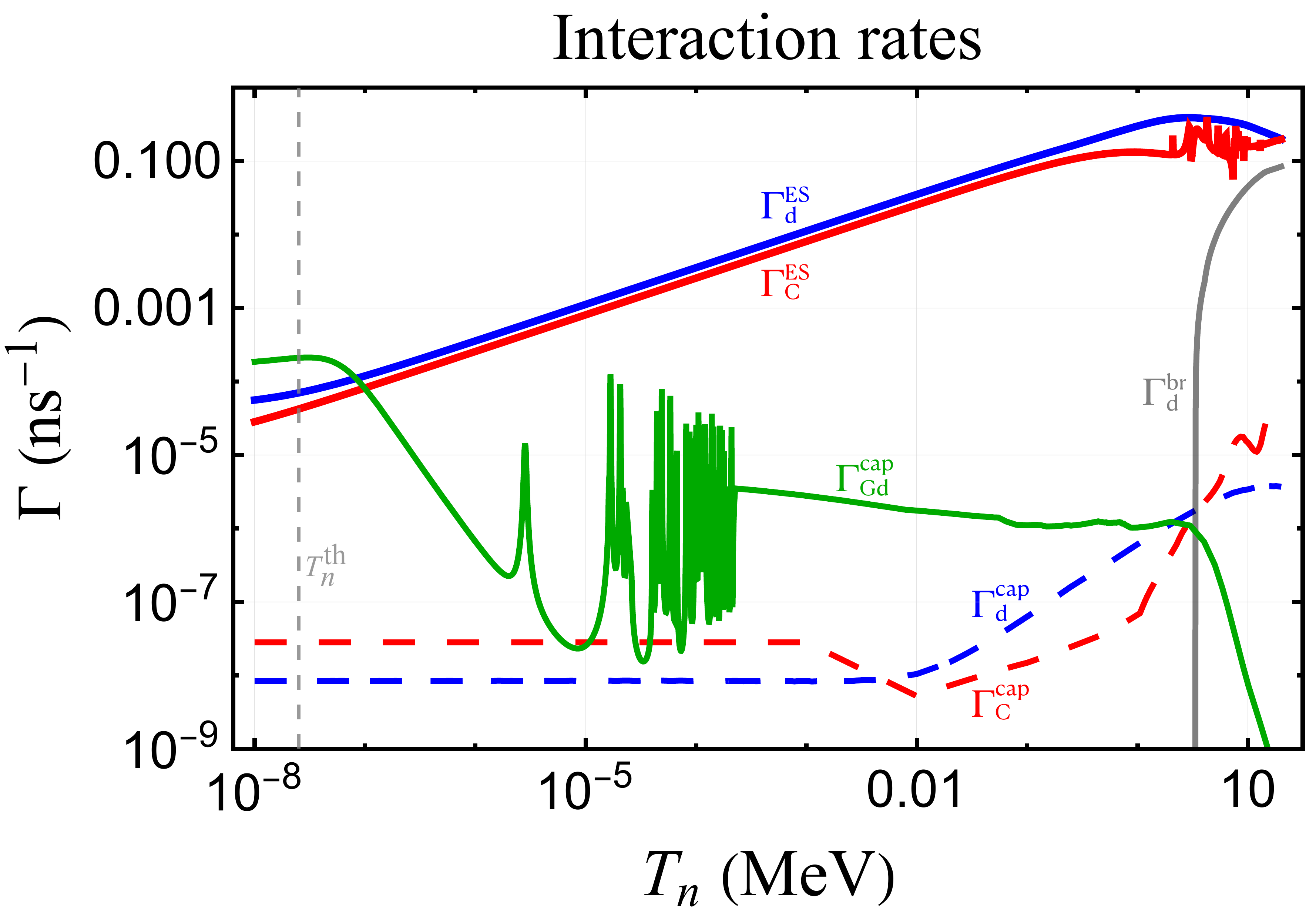}
		\caption{\label{fig:int} The relevant interaction rates ($\Gamma = n_x \sigma_{nx} v_n$) of the 
			neutron are shown. The elastic scattering off deuteron (blue) and carbon 
			(red) are indicated with solid lines. The neutron capture on deuteron (blue) 
			and carbon (red) are shown by dashed lines. The green curve shows the interaction rate for 
			neutron capture on 0.1\% mass concentration of $^{157}$Gd. The gray curve corresponds 
			to the neutron induced breakup 
			of the deuteron. We also indicate the thermal neutron energy ($T_n^{\rm th} = 0.025$eV) 
			by 
			the vertical dashed line. }
	\end{figure}

	A recoil neutron produced through neutrino dissociation of the deuteron interacts with the 
	neighboring deuteron, carbon, and gadolinium nuclei. The prominent interactions include the 
	elastic scattering, capture, and deuteron break-up. A simple way to examine these interactions 
	in 
	the detector is to compare the interaction rates, $\Gamma = n_x \sigma_{nx} v_n$. Since the 
	neutrons are non-relativistic, we use $v_n \sim 1.4~\text{cm/ns}~(T_n/\text{MeV})^{1/2}$. The 
	number densities of target are adapted from Appendix \ref{estNd} and the cross sections are 
	taken from EXFOR database\footnote{\url{www-nds.iaea.org/exfor/endf.htm}}. These interaction 
	rates are shown in Figure \ref{fig:int}. \\
	
	As can be inferred from Figure \ref{fig:int}, interactions of the \emph{fast} neutrons ($T_n > 1$ 
	MeV) are 
	dominated by elastic collisions off deuteron and carbon. On an average, such neutrons loses 
	 4/9th ($\sim 44$\%) of its kinetic energy in a neutron-deuteron scattering, and 24/169th 
	 ($\sim 14\%$) in 
	neutron-carbon scatterings. As a result, the recoiling neutrons slow down, thermalise, and 
	eventually get 
	captured producing detectable $\gamma$ photons. Conservatively assuming only interactions 
	with deuteron, the number of elastic scatterings required for thermalisation is approximated by,
	\begin{equation}
		N_{\text{ES}} = \frac{1}{\log\left(5/9\right)} \log\left( \frac{T_n^{\rm th}}{T_n}	\right)      \approx 
		4~\log_{10}\left( \frac{T_n}{T_n^{\rm th}}	\right).
	\end{equation}
	where $T_n^{\rm th} = 0.025$ eV ($\sim 300$ K) is the typical thermal energy of a neutron. For 
	example, neutron with initial kinetic energy 1 (6) MeV undergoes approximately 30 (33) elastic 
	scatterings before thermalisation. Since the energy loss rate on $\ce{^{12}C}$ is smaller than 
	deuteron, inclusion of carbon interactions leads to slightly slower thermalisation. \\

	In a 100\% deuterated scintillator without Gd, the thermal neutron 
	can be captured on deuteron $(n+d\rightarrow \ce{^3H} + \gamma)$ or carbon $(n + 
	\ce{^{12}C} \rightarrow \ce{^{13}C} + \gamma)$. The cross 
	section for capture on deuteron is 0.5 mb i.e., $0.5 
	\times 10^{-27}$ cm$^2$ whereas for $\ce{^{12}C}$, the cross section is 3.5 mb i.e., $3.5 
	\times 
	10^{-27}$ cm$^2$. Using the 
	effective elastic scattering rate ($\Gamma^{\text{ES}} = \Gamma^{\text{ES}}_d + 
	\Gamma^{\text{ES}}_\text{C}$) and 
	effective capture rate ($\Gamma^{\text{cap}} = \Gamma^{\text{cap}}_d + 
	\Gamma^{\text{cap}}_\text{C}$)
	we estimate that the thermal neutron undergoes 
	\begin{equation}
		N_{\text{cap}}^{\text{no Gd}} = 
		\frac{\Gamma^{\text{ES}}(T_n^{\rm th})}{\Gamma^{\text{cap}}(T_n^{\rm th})} \approx 
		3000
	\end{equation} 
	scatterings off deuteron and carbon before getting captured. This results in a delay of 
	$\mathcal{O}$(20 ms) between the prompt scintillation and the neutron capture signal. 
	Approximating this as a random walk, we obtain the diffusion length of the thermal neutron to be 
	174 cm. Such a long time interval and large travel distance from the prompt scintillation signal 
	will make tagging neutrons difficult. This issue of long neutron capture time in 
	Super-Kamiokande is resolved by adding Gd \cite{Beacom:2003nk}. The cross section for 
	capture 
	of thermal neutron  
	on $\ce{^{155}Gd}$ and $\ce{^{157}Gd}$ is $6.11\times10^{-20}\text{ 
		cm}^2 ~\text{and } 
	2.54\times10^{-19}\text{ cm}^2$ respectively\footnote{The 
		natural abundance of $\ce{^{155}Gd}$ and $\ce{^{157}Gd}$ are 14.8\% and 15.6\% 
		respectively.}. As a consequence, for 0.1\% mass concentration of Gd ($\sim10^{-4}$ by 
		number 
	density), the thermal neutrons are almost immediately captured and the delay is reduced to 
	$\mathcal{O}(50)~\mu$s. Thus, addition of Gd reduces the neutron loss rate and greatly 
	enhances the tagging efficiency. \\
	
		Assuming partial deuteration results in a significant number of free protons in the DLS. The 
	larger elastic scattering cross section would imply even quicker thermalisation of neutrons, but 
	the free protons introduces an additional problem of segregating  
	elastic scattering on proton ($\nu,\bar{\nu} + p \rightarrow \nu,\bar{\nu} + p$) from the NC 
	dissociation ($\nu,\bar{\nu} + d \rightarrow \nu,\bar{\nu} + p + n$) where the neutron is 
	not detected. In addition, having lesser target deuterons would reduce the expected 
	signal as well.\\ 
	
	\subsection{Secondary breakups}
	
\begin{figure}[h!]
		\centering
		\includegraphics[width=7cm]{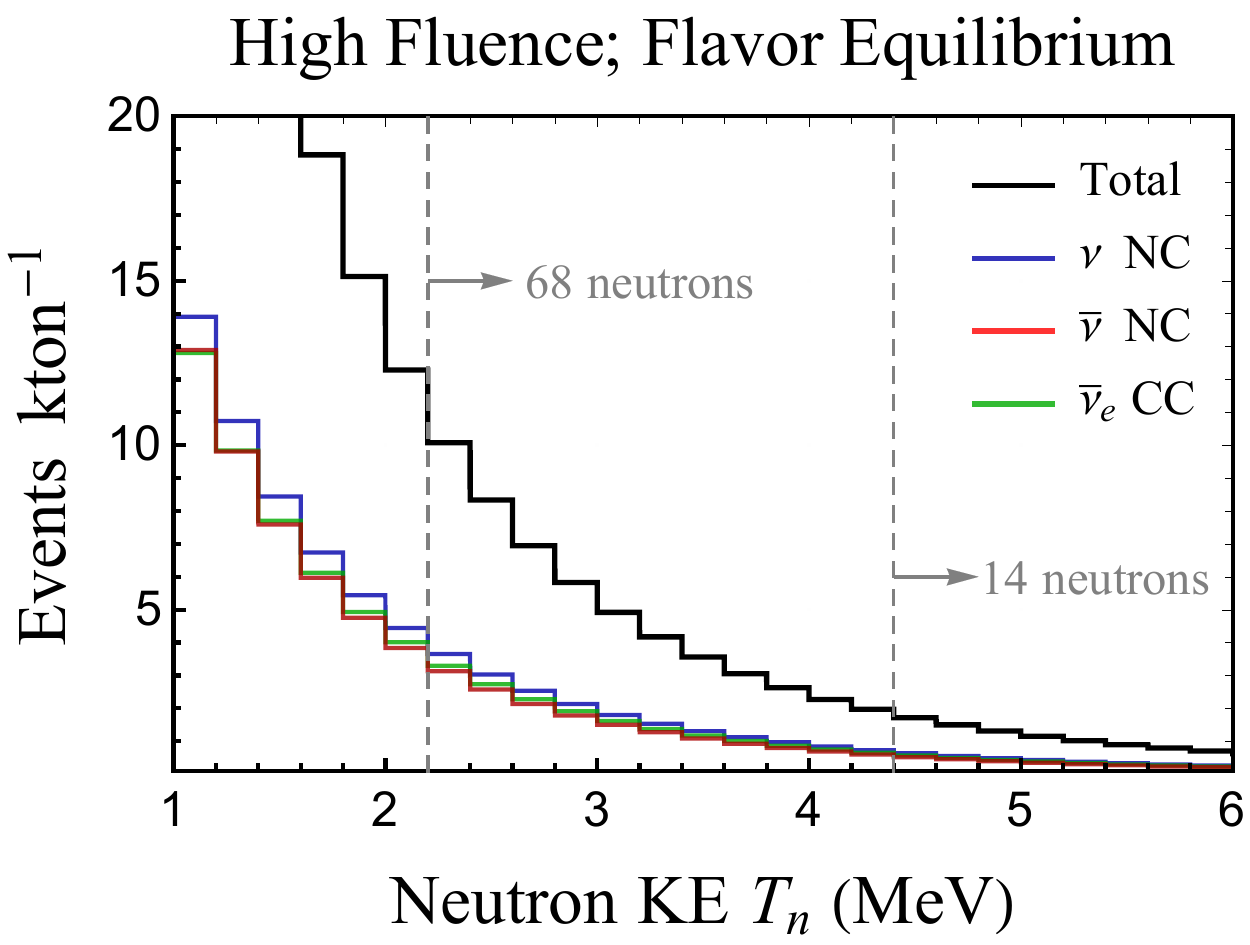}
		\includegraphics[width=7cm]{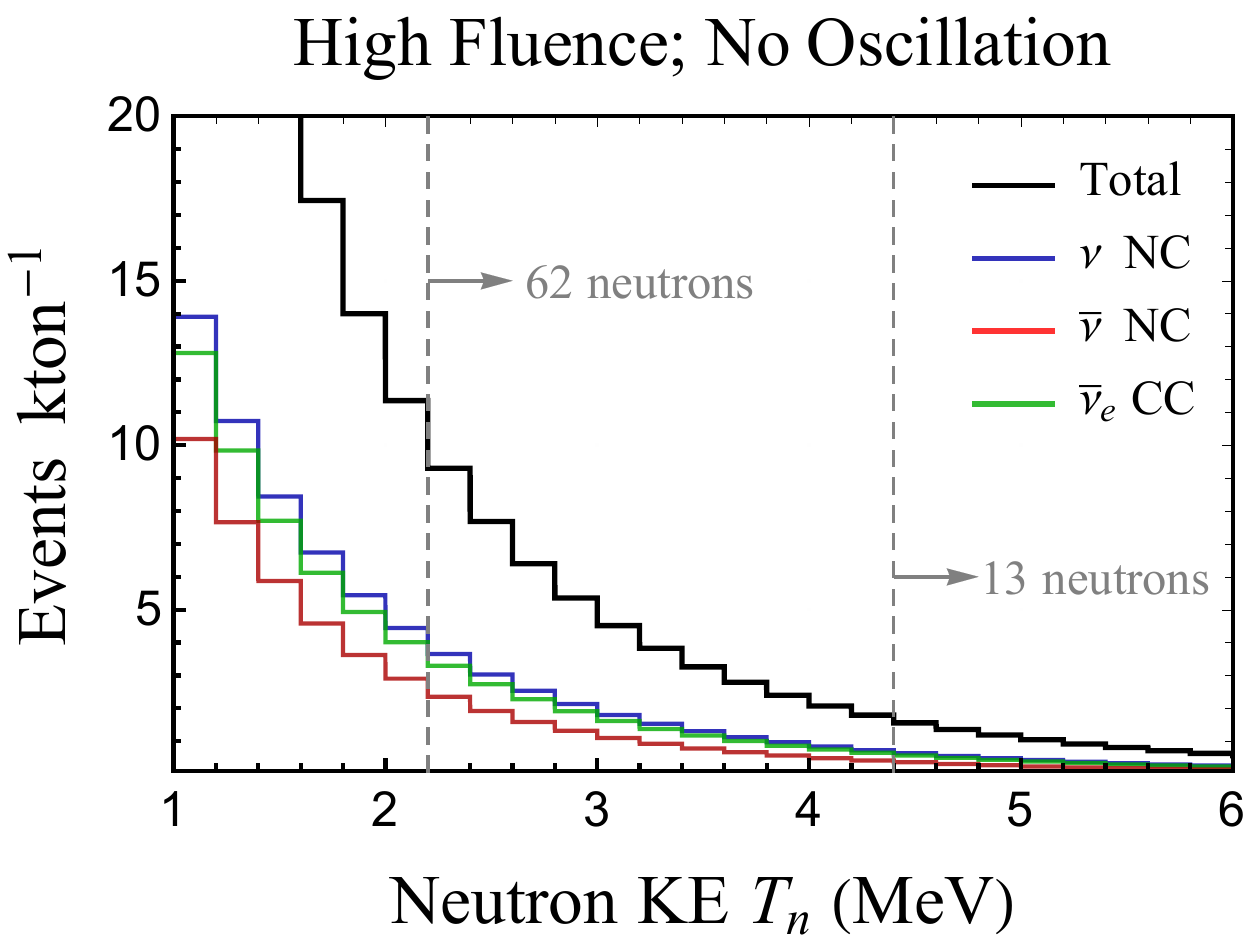} \\[0.5 cm]
		\includegraphics[width=7cm]{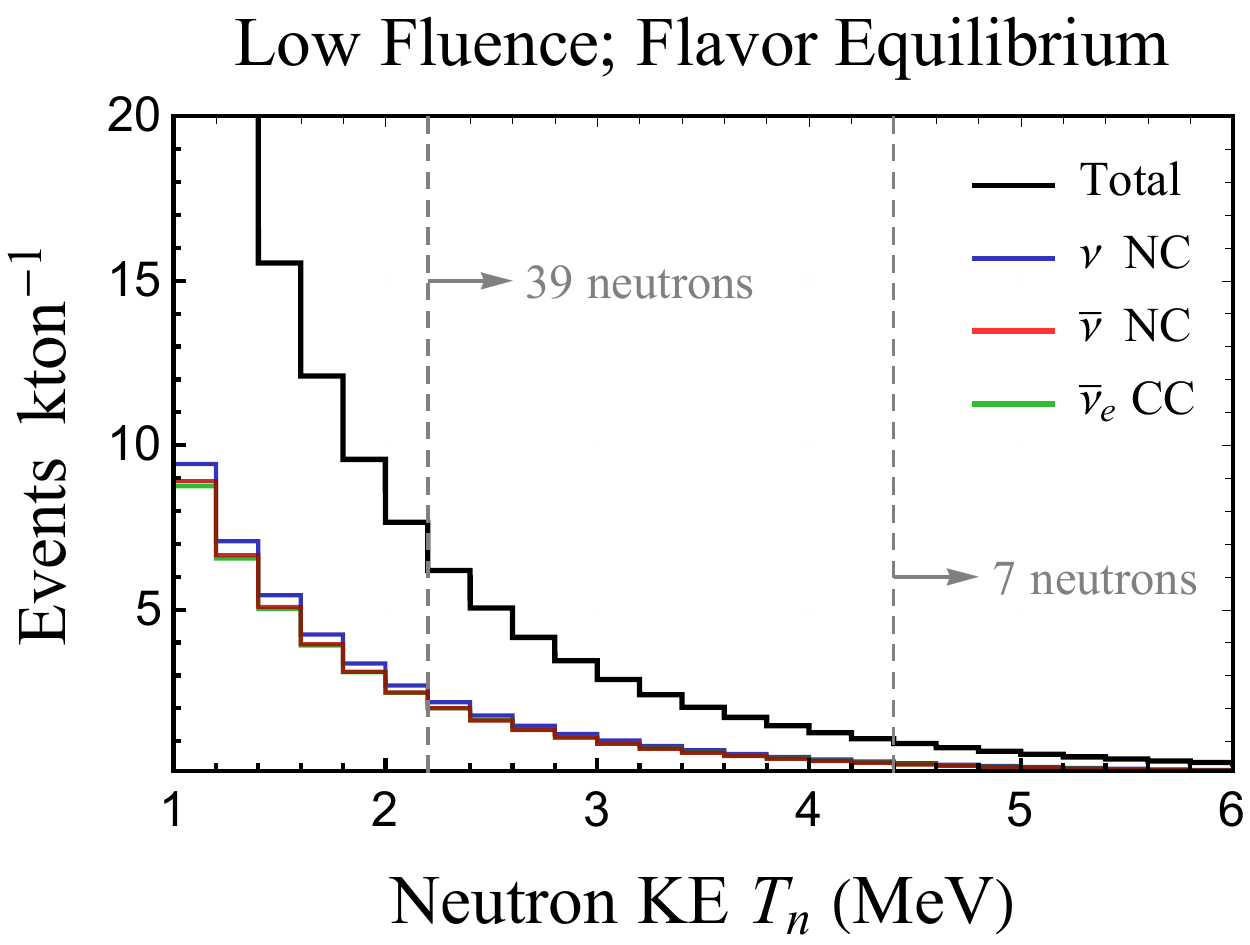}
		\includegraphics[width=7cm]{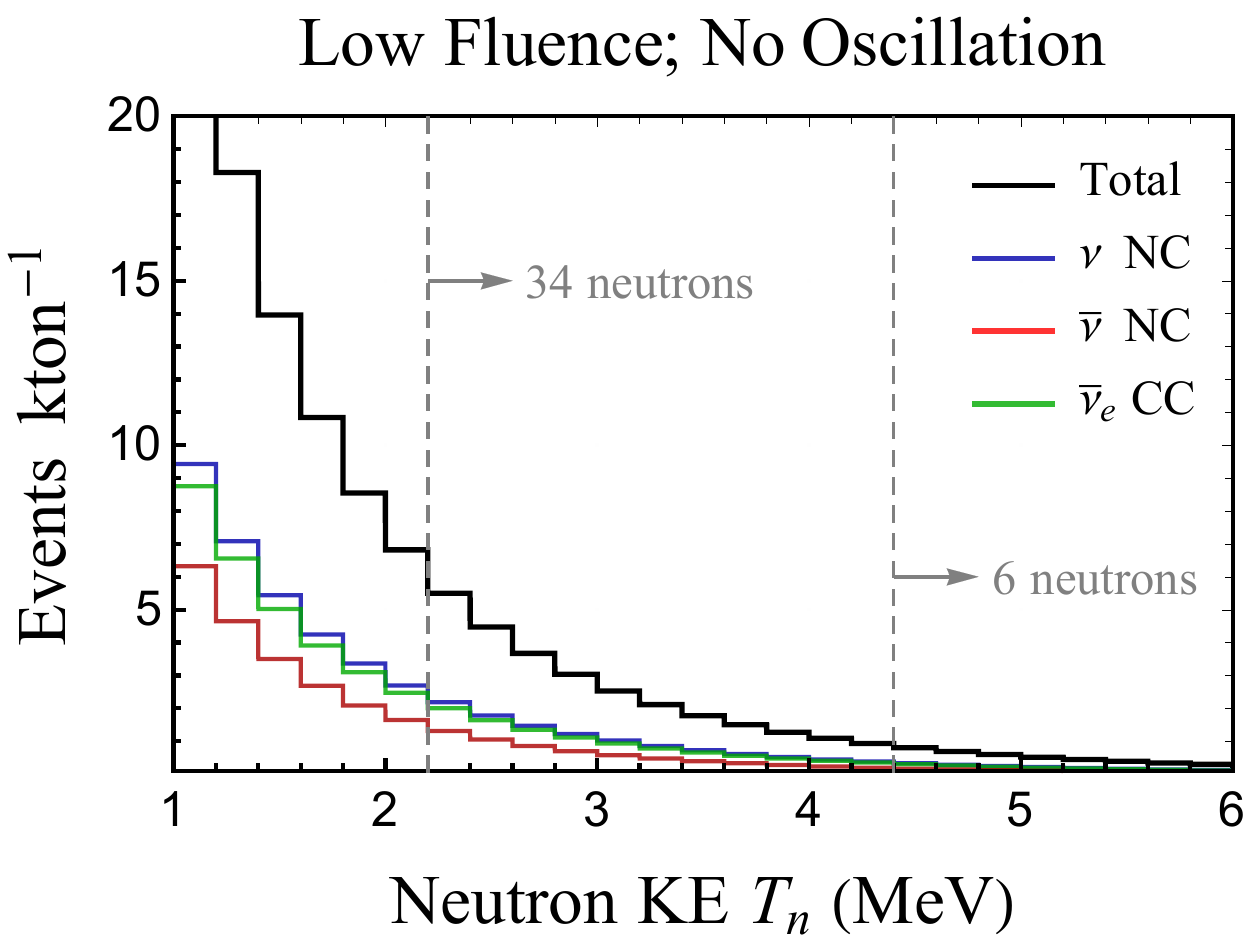}
		\caption{\label{fig:neutron} The spectrum of events with respect to the kinetic energy 
			of the recoiling neutron is shown for choices of fluence parameters (High and Low) and 
			flavor 
			conversion scenario (flavor equilibrium and no oscillation) considered in this paper. We have 
			shown the count of neutrons having kinetic energy above the deuteron breakup 
			threshold as well as the count above twice the threshold.}
	\end{figure}

	The recoiling neutrons having kinetic energy larger than deuteron binding energy can, in 
	principle, dissociate the deuteron. In Figure \ref{fig:neutron}, we have shown the distribution of 
	recoiling neutron kinetic energy for various scenarios considered in this paper. Depending on the 
	fluence model and flavor conversion scenario, one expect 35-70 neutrons ($\sim$10\% of total 
	event rate) that are above the breakup threshold. These secondary breakups cannot 
	be distinguished from neutrino induced dissociation and would be considered as 
	\emph{background} during reconstruction. \\ 
	
\begin{figure}[h!]
		\centering
		\includegraphics[width=9cm]{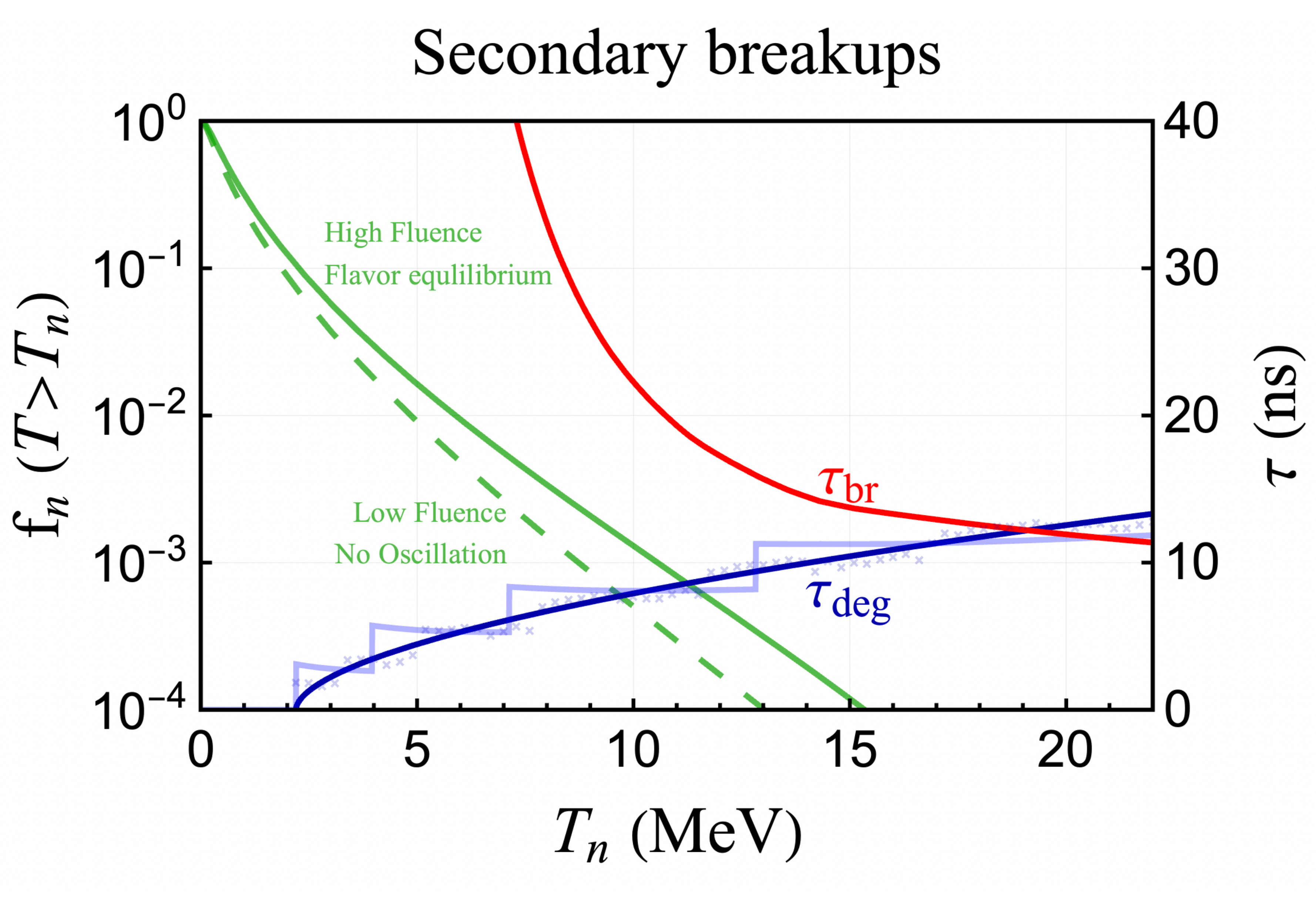}
		\caption{\label{fig:deg} The interaction time for breakup ($\tau_{\text{br}}$, red) and 
		degradation 
		($\tau_{\text{deg}}$, blue) are shown with ordinate on the right. The degradation time, 
		evaluated numerically by including scattering with carbon, is shown with blue crosses. The 
		darker blue curve is a fit to the numerical data. The lighter blue curve shows 
		the degradation time obtained 
		from Eq.\eqref{eq:deg} which only considers scattering with deuterons. We also 
		show the fraction of neutrons with energy greater than $T_n$ (green) on a log scale with 
		ordinate on the left. One can 
		infer that the two time scales are comparable for $T_n > 15$ MeV, but the fraction of  
		neutrons at these energies is negligible. }
	\end{figure}

	From Figure \ref{fig:int}, it is clear that the interaction rate for breakup is less than that of elastic 
	scattering. To compare the two interactions, we look at two time scales, (i) the time for breakup 
	($\tau_{\text{br}}\sim 1/ \Gamma_{d}^{\text{br} }$), and (ii) the degradation time 
	($\tau_{\text{deg}}$) i.e. the time required for a neutron to lose energy via elastic scatterings 
	and go below the breakup threshold. Considering only deuteron targets, the number of 
	scatterings required for degradation is, 
	\begin{equation}
		N_{\text{deg}} \approx 4~\log_{10}\left( \frac{T_n}{2.2~\text{MeV}}	\right) \Theta(T_n - 
		2.2~\text{MeV})
	\end{equation}
	where $\Theta$ is the Heaviside step function. The degradation time can be estimated by a sum 
	over interaction time for $N_{\text{deg}}$ scatterings and given as,  
	\begin{equation}
		\label{eq:deg}
		\tau_{\text{deg}} = \frac{1}{n_d }\sum_{i=1}^{ \lceil N_{\text{deg}}\rceil}\frac{1}{ \sigma_i v_i} 
		\approx \frac{1}{n_d 
			\langle \sigma \rangle} \sqrt{ \frac{M_n}{2 T_n} }  \sum_{i=1}^{\lceil N_{\text{deg}}\rceil} \left( 
			\frac{9}{5} 
		\right)^{i/2}
	\end{equation}
	where $\sigma_i$ denotes the elastic scattering cross section at $(5/9)^i \times T_n$ and 
	$\langle 
	\sigma 
	\rangle = 2\times10^{-24}\text{ cm}^2$ is the average cross section at these energies. The 
	results are shown in Figure 
	\ref{fig:deg} from which one can conclude that the two interactions are comparable for 
	$T_n>15$ MeV. In the same figure, we also show the fraction of neutrons having energy greater 
	than $T_n$. We find that, only a negligible fraction of the recoil neutrons would participate in 
	secondary breakups. In conclusion, almost all of the recoil neutrons would lose energy through 
	elastic collisions and secondary breakups can be ignored. So far, we have not included elastic 
	scattering on carbon nuclei for brevity. We include these interactions numerically and find that 
	the degradation time is similar at these energies and our conservative conclusion is unchanged.
	
	\subsection{Secondary scintillation}
	
\begin{figure}[h!]
		\centering
		\includegraphics[width=7cm]{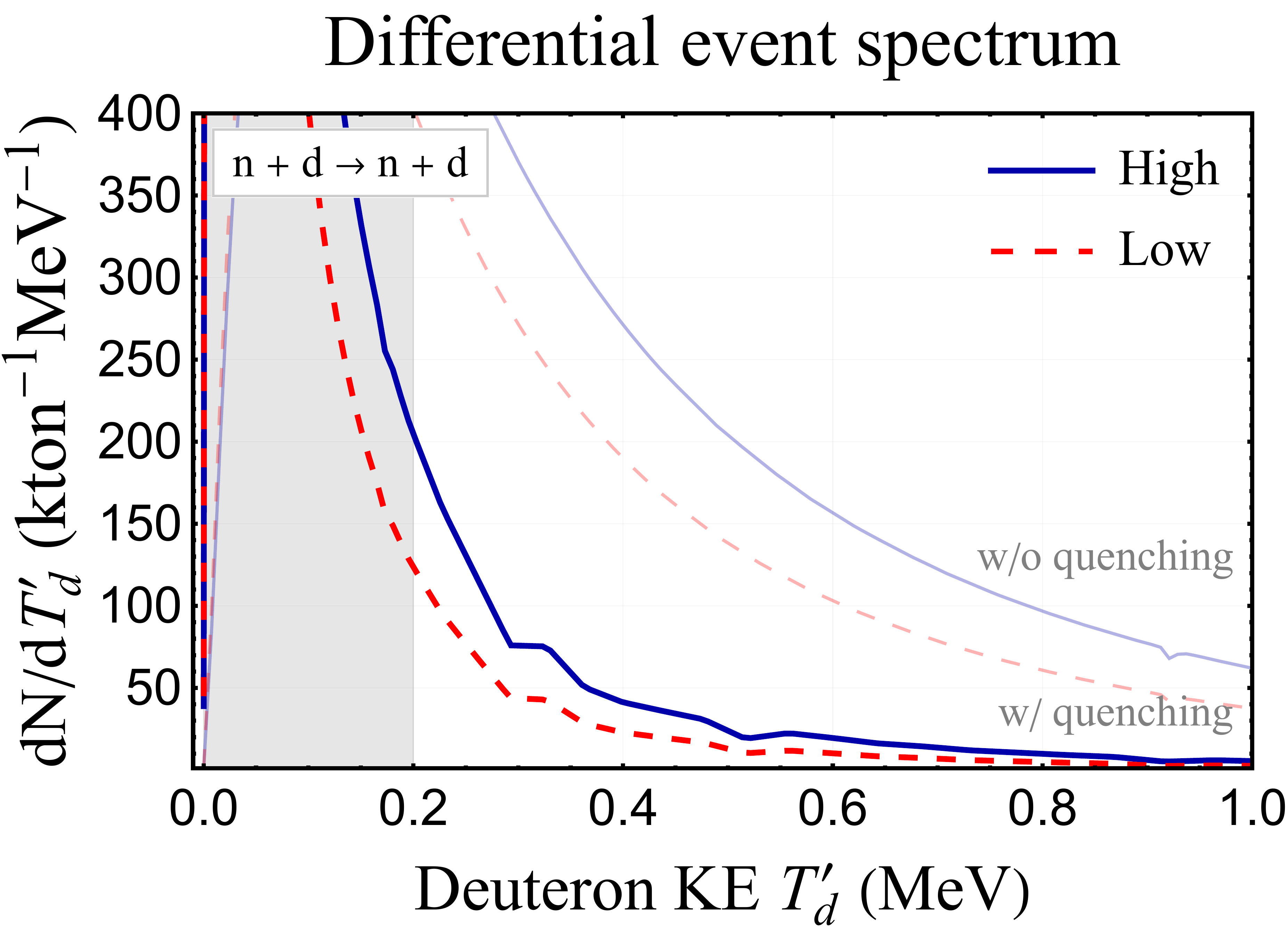}\hspace{.5cm}
		\includegraphics[width=6.9cm]{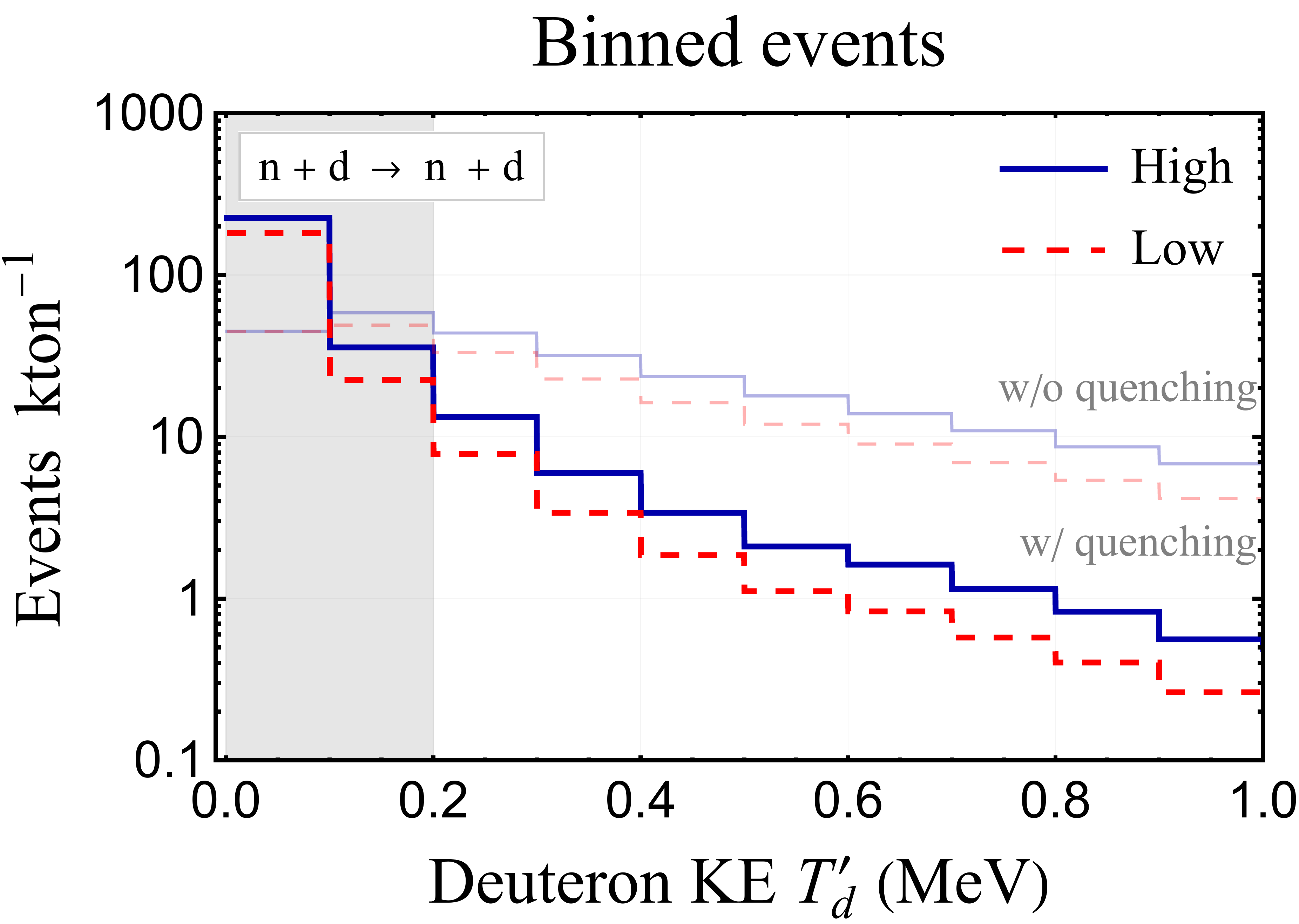}
		\caption{\label{fig:deuteron}  The differential event spectrum (left) and binned events (right) 
			with visible deuteron recoil energy with and without quenching are shown. The solid blue 
			curves are for optimistic fluence model High and dashed red curves are for the 
			conservative fluence model Low. The gray shading shows energies below threshold, where 
			we take $E_{th}$ = 200 keV due to $^{14}$C beta decay background. Note that these 
			events have a proton scintillation signal and a neutron capture signal available for tagging.}
	\end{figure}

	The elastic scattering between recoiling neutrons and deuteron in the detector leads to a 
	secondary scintillation signal from the recoiling deuteron, typically separated by a few ns. Almost 
	always, it's the first scattering of the neutron that would result in visible signal. In CC 
	interactions, 
	the scintillation from the positron will overwhelm the scintillation from the primary protons as well 
	as the secondary deuterons. In NC interactions, a detector with very good 
	timing resolution\footnote{for example, LAPPDs (Large 
	Area Picosecond Photo-Detectors) 
			\cite{Lyashenko:2019tdj}} can, in principle, separate the primary and secondary scintillation 
			signals. To 
	estimate this we look at the event spectrum,
	\begin{equation}
		\frac{dN}{dT_d} = \frac{\Gamma^{\text{ES}}_d}{\Gamma^{\text{ES}}} \left( \frac94 
		\right) \frac{dN}{dT_n}
	\end{equation}
	where the relative interaction rate accounts for the fraction of recoil neutrons scattering off 
	deuteron and $4/9~T_n$ is the average energy transfer to the deuteron. We only look at the 
	neutrons produced in NC interactions, hence the result is independent of the oscillation 
	scenarios and only depends on the fluence model. The deuteron scintillation will be quenched, 
	similar to proton. Assuming the Bethe-Bloch formula, the energy loss rate for deuteron will be 
	approximately half of that of proton. The quenching factor is then approximately given by the 
	$k_B = 0.007$ cm MeV$^{-1}$ curve in Figure \ref{fig:quenching}. In Figure \ref{fig:deuteron}, 
	we have shown the distribution of these secondary scintillation events with respect to the visible 
	deuteron energy $T_d^\prime$. We find that approximately 31 (17) of these secondary 
	scintillation signals are above threshold for the High (Low) fluence model. One must note that, 
	these secondary scintillations are preceded by proton scintillation signal, and followed by a 
	neutron capture signal. Moreover, the secondary signals correspond to the high energy part of 
	the proton signal, and if they cannot be separated they will lead to deterioration in the expected 
	energy resolution at higher energies.
	
	\section{Summary} \label{conc}
	
	In this paper, we have shown the potential of a deuterated liquid scintillator based detector for 
	observing supernova neutrinos. We employed the framework of $\slashed{\pi}$EFT to evaluate 
	the 
	cross sections for neutrino dissociation of the deuteron.  We checked that the evaluated cross 
	sections are appropriate for supernova neutrino energies and have computed the event rates in 
	1 
	kton of DLS. We find that for an optimistic choice of supernova neutrino parameters, one can 
	observe up to 435 NC events, 170 $\nu_e$ and 108 $\bar\nu_e$ CC events for a typical galactic 
	supernova at a distance of 10 kpc from Earth. This ability to detect all flavors in a single detector 
	can allow a model-independent determination of the flavor-mixing 
	scenario~\cite{Capozzi:2018rzl}.\\
	
	We have also looked at the spectrum of events for these interaction channels. For NC, we 
	looked 
	at 
	the quenched scintillation signal from the proton in the final state. We show that, for threshold at 
	200 keV, one can observe up to 85 scintillation events from the proton. The spectral analysis of 
	these events can, in principle, be used to reconstruct the incident neutrino spectrum. 
	Observation 
	of these NC events will allow for independent normalization of the supernova neutrino flux as 
	well 
	as aid in understanding intricate oscillation physics. \\
	
	Most of the operational neutrino detectors rely on IBD to detect $\overline{\nu}_e$. A DLS is not 
	only sensitive to the elusive $\nu_e$, but the cross section for this channel is larger that the 
	other channels. Assuming flavor equilibrium, one obtains up to 170 events for $\nu_e$ and up to 
	110 events for $\overline{\nu}_e$. These two CC interaction channels have characteristic 
	signature in the detector due to different final state particles. We have also shown the spectrum 
	of these events which can also be used to reconstruct the incident neutrino spectrum. \\
	
	We have also studied the interactions of recoiling neutrons in DLS. Comparing their interaction 
	rates, we see that the neutrons elastically scatter off the deuteron and carbon nuclei,  
	thermalise with the medium, and eventually get captured. While the breakup of deuteron by 
	neutron 
	can be safely ignored, the recoil deuterons can lead to a quenched but visible scintillation signal. 
	For CC 
	interactions, such secondary scintillations will be difficult to separate from the positron signal. 
	For NC interactions, depending on the detector sensitivity, these signals can be used to tag the 
	primary proton signal, or lead to deterioration of resolution at high energies. These events are 
	followed by a neutron capture which can be tagged to get rid of the backgrounds. \\
	
	The $\gamma$s from the neutron capture are a prevalent signature in all of the interaction 
	channels except the CC interaction of $\nu_e$. This allows reduction of backgrounds \emph{if} 
	the 
	neutron can be tagged. Here we assumed that the thermal neutron always gets captured, which 
	is 
	an excellent assumption in presence of Gd. Without Gd, the neutrons have a large diffusion 
	length and might escape the detector. The neutron capture efficiency is also enhanced, if 
	the detector will have some fraction of free protons due to partial deuteration of the 
	scintillator. However, this introduces an additional complexity of segregating $\nu-p$ elastic 
	scattering with $\nu-d$ NC dissociation where the recoiling neutron escapes undetected. 
	Depending on the technological challenges with manufacture of the Gd doped DLS, as well as its 
	physics performance, suitable optimization is needed, keeping various constraints in mind. \\
	
	Reconstruction of the neutrino spectrum using the NC events in a DLS based detector will be 
	challenging but important. The key challenge concerns not the 
	quenching~\cite{Dasgupta:2011wg}, but rather the absence of one-to-one correspondence 
	between the  incident neutrino energy and the measured proton energy in a three body final 
	state. Nevertheless, the differential scattering cross section is known and can be inverted, as in 
	Ref.\,\cite{Dasgupta:2011wg}. An unfolding based approach, including the detector systematics, 
	may be useful~\cite{, Lu:2016ipr, Li:2017dbg, Li:2019qxi, Nagakura:2020bbw}.\\
	
	A DLS can also be used to study other aspects of low energy neutrino physics. Our preliminary 
	estimates indicate that a kton of DLS will not be sufficient to observe the diffuse supernova 
	background, given existing constraints from Super-K~\cite{Zhang:2013tua}. It has promising 
	applications in solar neutrino physics  where the $\nu_e$ can be detected at low 
	energies~\cite{dls:white}. Moreover, such a detector can also probe beyond standard model 
	physics such as proton decay \cite{Miura:2016krn, TheKamLAND-Zen:2015eva, Ahmed:2003sy}, 
	neutrinoless double beta decay \cite{Murayama:2003ci, Paton:2019kgy}, and axions from the 
	Sun 
	\cite{Bhusal:2020bvx}.\\
	
	The observation of the next galactic supernova promises to significantly improve our 
	understanding of core-collapse supernovae and the nature of its neutrino emission. However, 
	a lot more work remains to be done in simulating supernovae and their neutrino fluxes and in 
	computing the nature of flavor conversions therein. Although we now have several large neutrino 
	detectors capable of detecting supernova neutrinos, substantial efforts are underway to develop 
	adequate detection and analysis 
	strategies~\cite{Lujan-Peschard:2014lta,GalloRosso:2017mdz,Capozzi:2018rzl,Li:2020ujl}. It is 
	our 
	understanding that a DLS based detector will greatly aid in this task.\\ 
	
	\acknowledgments
	We thank J.W.\,Chen, for a useful discussion on the $B_0$ integral and its implementation. We 
	especially thank John Beacom, Amol Dighe, Milind Diwan, Ranjan Laha, M.V.N. Murthy, and R. 
	Palit for valuable comments and suggestions. We also thank the anonymous referee for 
	suggesting investigation into the secondary interactions of the recoiling neutron which forms 
	Section \ref{sec:Secondary} of our paper. 
	 The work of BD is supported by the Dept.\,\,of 
	Atomic Energy (Govt.\,\,of India) research project under Project Identification No. RTI 4002, the 
	Dept.\,\,of Science and Technology (Govt.\,\,of India) through a Swarnajayanti Fellowship, and by 
	the Max-Planck-Gesellschaft through a Max Planck Partner Group.
	
	\appendix
	
	\section{Revisiting neutrino deuteron interactions} \label{revist}
	
	The cross section for neutrino dissociation of deuteron can be calculated in two independent 
	ways. 
	In the Phenomenological Lagrangian Approach (PhLA), the nuclear transition matrix elements 
	are 
	obtained using the hadronic currents and the nuclear wave functions. The hadronic currents are 
	obtained with one body currents using impulse approximation and two-body meson exchange 
	currents. The initial and final state nuclear wave functions are obtained by solving the 
	Schr\"{o}dinger equation for phenomenological nucleon-nucleon (NN) potential. In the paper by 
	Nakamura, Sato, Gudkov, and Kubodera (NSGK) \cite{Nakamura:2000vp}, the authors used the 
	Argonne-$v$18 NN potential to obtain the cross section. On the other hand, the cross section 
	can 
	also be calculated in the framework of Pionless Effective Field Theory ($\slashed{\pi}$EFT) valid 
	for 
	$E_\nu \ll m_\pi$ \cite{Chen:1999tn}. In this approach, the massive hadronic degrees of freedom 
	(for example $\pi$ 
	and $\Delta$) are considered non-dynamical and integrated out. The nuclear interactions are 
	then 
	calculated perturbatively using the ratio of light and heavy scales as the expansion parameter. 
	The 
	light scales in the theory are determined by the nuclear scattering lengths ($\sim 10$ MeV) and 
	the 
	deuteron binding momentum ($\sim50$ MeV) whereas the heavy scale (also the cut-off) is given 
	by the pion mass ($m_\pi \sim 100$ MeV). In the paper by Butler, Chen, and Kong (BCK) 
	\cite{Butler:2000zp}, the authors have obtained neutrino-deuteron cross sections at 
	next-to-next-to-leading-order. \\
	
	The total cross sections obtained by both methods are in excellent agreement. In PhLA, there 
	are 
	many undetermined form factors whose numerical values are obtained by matching the theory 
	predictions to other experiments (for example, photo-dissociation of deuteron, i.e., $d + \gamma 
	\rightarrow n + p$). As the parameters are determined by precise experiments, the predictions of 
	neutrino-deuteron cross section in this approach are very robust. Whereas in $\slashed{\pi}$EFT, 
	there are no undetermined parameters (barring the counter-term $L_{1,A}$), as everything can 
	be 
	systematically obtained starting from Standard Model. On the flip side, PhLA is valid up to very 
	large neutrino energies ($>150$ MeV) but $\slashed{\pi}$EFT starts to become inaccurate at 
	few 
	tens of MeV. In Figure\,\ref{comp} we see that the total cross sections obtained by NSGK and 
	BCK 
	are very similar up to neutrino energy of 50 MeV. Thus we can safely use $\slashed{\pi}$EFT for 
	estimating the supernova neutrino event rates.\\
	
	The two approaches are independent and complementary. In order to match the total cross 
	sections with NSGK at a percent level accuracy, the numerical value of the unknown counter 
	term 
	in BCK is obtained in what would be the \emph{natural} range for the parameter. The agreement 
	of 
	these approaches at low energy also serves to validate the approximations made in PhLA. \\

	In the trailing discussion, we outline the relevant results from BCK \cite{Butler:2000zp} to 
	correct 
	the above said typo. The differential cross section for neutrino-deuteron interaction is given by
	\begin{equation}
		\frac{d^2 \sigma}{dE^\prime d\Omega} = \frac{G_F^2}{32 \pi^2} \frac{ \mathbf{k^\prime} }{ 
			\mathbf{k} }S_1( \mathbf{k^\prime} ) \ell_{\mu \nu} W^{\mu \nu}
	\end{equation}
	where $\mathbf{k}~(\mathbf{k^{\prime}})$ is the three momentum and $E~(E^\prime)$ is the 
	energy of the initial (final) lepton. The Coulomb interaction between final state particles is 
	encapsulated by $S_1$ which is significant at low energies and thus relevant for supernova 
	neutrino interactions. The lepton current tensor is given by, 
	\begin{equation}
		\ell^{\mu \nu} = 8 \left( k^\mu k^{\prime \nu} + k^\nu k^{\prime \mu} - k\cdot k^\prime g^{\mu 
			\nu} \pm i \varepsilon^{\mu \nu \rho \sigma} k_\rho k^\prime_\sigma  \right)
	\end{equation}
	and the hadronic current tensor ($W_{\mu \nu}$) is given in terms of six independent structure 
	functions. Of these, only three ($W_1, ~W_2, ~W_3$) give non-zero contribution. In terms of 
	these, the differential cross section is simplified as,  
	\begin{equation}
		\label{eq:diff}
		\frac{d^2 \sigma}{dE^\prime d\Omega} = \frac{G_F^2 E_\nu  \mathbf{k^{\prime}} }{2 \pi^2} 
		S_1( 
		\mathbf{k^\prime} ) \left[ 2 W_1 \sin^2 \frac\theta 2  + W_2 \cos^2 \frac \theta 2  \mp 2 
		\frac{(E_\nu + E^\prime)}{M_d} W_3 \sin^2 \frac \theta 2 \right]
	\end{equation}
	where the - (+) sign in front of the third term is for (anti-) neutrino in the initial state.\\ 
	
	We use the perturbative expressions for the structure functions $W_{1,2,3}$, as given in 
	BCK~\cite{Butler:2000zp}.  However, Eq.(53) in BCK, which includes the Coulomb interaction 
	between the final state protons in the CC interaction $\nu_e + d \to e + p + p$, has a typo, 
	where 
	$k^2$ in the denominator of the last term was mistakenly printed as $q^2$. The correct 
	expression is 
	\begin{equation}
		B_0 = M_N \int \frac{d^3k}{(2\pi)^3} \frac{\sqrt{8 \pi \gamma}}{k^2 + \gamma^2} \frac{2 \pi 
			\eta_k}{e^{2 \pi \eta_k} - 1} \frac{e^{2 \eta_k \tan^{-1}(k/\gamma)}}{p^2 - k^2 + i \epsilon} 
		\left[ 1 
		+ \frac{\textbf{q}^2 \left( ( 1 - 2 \eta_k^2) k^2 - 6 \eta_k  k \gamma - 3 \gamma^2 
			\right)}{12({k}^2 + \gamma^2)^2}   \right]
	\end{equation}  
	where $M_N = 939$\,MeV is the mass of a nucleon, $\gamma = 45.70$\,MeV is the deuteron 
	internal momentum, and $\eta_k = \alpha M_N/(2k)$. The fine structure constant $\alpha = 
	1/137$ 
	at these scales. The magnitude of the relative momentum between the protons is $2p$ and 
	$\mathbf{q}$ is the spatial component of the momentum transfer. Substituting for $\eta_k$, the 
	expression can be written as 
	\begin{align}
		B_0 &= \frac{M_N\gamma^{1/2}}{2 \pi^2} \int_{0}^{\infty} dk \frac{k^2}{k^2 + \gamma^2}  
		\frac{C[k]}{p^2 - k^2} \left[ 1 +  \frac{\textbf{q}^2 \left( 2 k^2 - M_N^2 \alpha^2 - 6 M_N 
		\alpha 
			\gamma - 6 \gamma^2\right)}{24(k^2 + \gamma^2)^2}\right]\,,
	\end{align}  
	where $C[k] = (2 \pi \eta_k) ( e^{2 \pi \eta_k} - 1)^{-1}e^{2 \eta_k \tan^{-1}(k/\gamma)}$ is 
	related 
	to the Coulomb interaction between the nucleons. This integral must be evaluated carefully: the 
	integrand has poles at $\pm i\gamma$ and $\pm p$, but $C[k]$ is undefined at $\pm i\gamma$; 
	so, one replaces $C[k] \to (C[k]-1)+1$, and the $k$-integral involving $C[k]-1$ is computed 
	numerically by taking the principal value at $k=p$, whereas the latter (for which the integrand is 
	even in $k$ and vanishes at $k\to\infty$) is computed using the method of residues. The 
	differential cross sections, thus obtained, are shown in Figure \ref{fig:diffxsNC} for NC and in 
	Figure 
	\ref{fig:diffxsCC} for CC interactions. 
	
	\newpage
	\section{Tabulated differential cross sections}
	\label{tabxsec}
	
	\subsection{$\nu + d \rightarrow \nu + n + p  $}
	\begin{table}[H]
		\centering
		\caption{\label{data_nu_nc} The differential cross section ($d\sigma/dT_p$ in units of 
			$10^{-42}$ cm$^2$ MeV$^{-1}$) for $\nu$ NC interaction is tabulated for various choices 
			of 
			recoiling proton kinetic energy $T_p$ and incoming neutrino energy $E_\nu$.}
		\begin{tabular}{c c c c c c c }
			\toprule 
			$T_p\text{ (MeV)}$ & $E_\nu = $ 5 MeV & 10 MeV & 15 MeV &20 MeV &30 MeV & 50 MeV 
			\\
			\midrule
			0.001 & 0.032 & 0.041 & 0.042 & 0.043 & 0.044  & 0.045 \\
			0.01 & 0.325 & 0.406 & 0.420 & 0.429 & 0.441  & 0.452 \\
			0.05 & 0.380 & 3.161 & 3.287 & 3.125 & 3.038 & 3.022 \\
			0.1 & 0.289 & 2.701 & 7.627 & 6.731 & 6.135  & 5.875 \\
			0.2 & 0.170 & 1.758 & 5.556 & 12.770 & 11.510  & 10.260 \\
			0.3 & 0.106 & 1.228 & 3.907 & 8.981 & 16.190  & 13.560 \\
			0.4 & 0.068 & 0.899 & 2.905 & 6.635 & 20.440  & 16.220 \\
			0.5 & 0.044 & 0.679 & 2.239 & 5.125 & 18.320  & 18.470 \\
			1.0 & 0.004 & 0.214 & 0.806 & 1.932 & 6.977  & 26.590 \\
			2.0 & - & 0.031 & 0.184 & 0.505 & 2.020& 15.830 \\
			3.0 & - & 0.003 & 0.054 & 0.180 & 0.813  & 6.638 \\
			4.0 & - & - & 0.016 & 0.072 & 0.384  & 3.256 \\
			5.0 & - & - & 0.003 & 0.030 & 0.199  & 1.791 \\
			6.0 & - & - & - & 0.012 & 0.109  & 1.066 \\
			7.0 & - & - & - & 0.004 & 0.061 & 0.670 \\
			8.0 & - & - & - & 0.001 & 0.034  & 0.439 \\
			9.0 & - & - & - & - & 0.019  & 0.296 \\
			10. & - & - & - & - & 0.010  & 0.204 \\
			11. & - & - & - & - & 0.004  & 0.142 \\
			12. & - & - & - & - & 0.002  & 0.100 \\
			13. & - & - & - & - & -  & 0.071 \\
			14. & - & - & - & - & -  & 0.050 \\
			15. & - & - & - & - & - & 0.035 \\
			16. & - & - & - & - & -  & 0.024 \\
			17. & - & - & - & - & -  & 0.017 \\
			18. & - & - & - & - & - & 0.011 \\
			19. & - & - & - & - & -  & 0.007 \\
			20. & - & - & - & - & -  & 0.004 \\
			21. & - & - & - & - & -  & 0.002 \\
			22. & - & - & - & - & -  & 0.001 \\
			23. & - & - & - & - & -  & - \\
			24. & - & - & - & - & -  & - \\
			25. & - & - & - & - & -  & - \\
			\bottomrule 
		\end{tabular}
	\end{table}
	
	\subsection{$\bar{\nu} + d \rightarrow \bar{\nu} + n + p  $}
	\begin{table}[H]
		\centering
		\caption{\label{data_nubar_nc} The differential cross section ($d\sigma/dT_p$ in units of 
			$10^{-42}$ cm$^2$ MeV$^{-1}$) for $\bar{\nu}$ NC interaction is tabulated for various 
			choices of recoiling proton kinetic energy $T_p$ and incoming neutrino energy $E_\nu$.}
		\begin{tabular}{c c c c c c c }
			\toprule
			$T_p\text{ (MeV)}$ & $E_\nu = $ 5 MeV & 10 MeV & 15 MeV &20 MeV &30 MeV & 50 MeV 
			\\
			\midrule
			0.001 & 0.032 & 0.040 & 0.042 & 0.043 & 0.044  & 0.045 \\
			0.01 & 0.317 & 0.400 & 0.415 & 0.426 & 0.439  & 0.451 \\
			0.05 & 0.372 & 3.000 & 3.138 & 3.011 & 2.960 & 2.975 \\
			0.1 & 0.282 & 2.561 & 7.026 & 6.271 & 5.822  & 5.684 \\
			0.2 & 0.167 & 1.667 & 5.103 & 11.340 & 10.400 & 9.583 \\
			0.3 & 0.104 & 1.165 & 3.590 & 7.980 & 14.030  & 12.240 \\
			0.4 & 0.067 & 0.853 & 2.669 & 5.897 & 17.090  & 14.170 \\
			0.5 & 0.043 & 0.644 & 2.057 & 4.555 & 15.170  & 15.650 \\
			1.0 & 0.004 & 0.203 & 0.740 & 1.715 & 5.776  & 19.890 \\
			2.0 & - & 0.030 & 0.169 & 0.447 & 1.668  & 11.310 \\
			3.0 & - & 0.003 & 0.049 & 0.159 & 0.670 & 4.750 \\
			4.0 & - & - & 0.014 & 0.064 & 0.317  & 2.334 \\
			5.0 & - & - & 0.003 & 0.027 & 0.164 & 1.287 \\
			6.0 & - & - & - & 0.010 & 0.090  & 0.768 \\
			7.0 & - & - & - & 0.003 & 0.051 & 0.485 \\
			8.0 & - & - & - & 0.001 & 0.028 & 0.319 \\
			9.0 & - & - & - & - & 0.016  & 0.216 \\
			10. & - & - & - & - & 0.008 & 0.150 \\
			11. & - & - & - & - & 0.004  & 0.105 \\
			12. & - & - & - & - & 0.001  & 0.075 \\
			13. & - & - & - & - & -  & 0.053 \\
			14. & - & - & - & - & -  & 0.038 \\
			15. & - & - & - & - & - & 0.027 \\
			16. & - & - & - & - & - & 0.019 \\
			17. & - & - & - & - & -  & 0.013 \\
			18. & - & - & - & - & -  & 0.009 \\
			19. & - & - & - & - & - & 0.005 \\
			20. & - & - & - & - & -  & 0.003 \\
			21. & - & - & - & - & -  & 0.002 \\
			22. & - & - & - & - & -  & 0.001 \\
			23. & - & - & - & - & - & - \\
			24. & - & - & - & - & -  & - \\
			25. & - & - & - & - & -  & - \\
			\bottomrule
		\end{tabular}
		
	\end{table}
	
	\subsection{$\nu_e + d \rightarrow e^- + p + p  $}
	\begin{table}[H]
		\centering
		\caption{\label{data} The differential cross section ($d\sigma/dE_e^\prime$ in units of 
			$10^{-42}$ cm$^2$ MeV$^{-1}$) for $\nu_e$ CC interaction is tabulated for various 
			choices 
			of 
			outgoing electron energy $E_e^\prime$ and incoming neutrino energy $E_\nu$. The 
			kinematic 
			cut-off $m_e \leq E_e^\prime \leq  E_\nu - 0.93$ MeV is apparent.  }
		\begin{tabular}{c c c c c c c }
			\toprule
			$E_e^\prime\text{ (MeV)}$ & $E_\nu = $ 5 MeV & 10 MeV & 15 MeV &20 MeV &30 MeV & 
			50 MeV \\
			\midrule
			0.5 & - & - & - & - & - & - \\
			1.0 & 0.004 & 0.001 & 0.000 & 0.000 & 0.000 & 0.000 \\
			2.0 & 0.034 & 0.003 & 0.001 & 0.000 & 0.000 & 0.000 \\
			3.0 & 0.169 & 0.011 & 0.003 & 0.001 & 0.000 & 0.000 \\
			4.0 & 0.049 & 0.029 & 0.007 & 0.003 & 0.001 & 0.000 \\
			5.0 & - & 0.070 & 0.013 & 0.005 & 0.002 & 0.000 \\
			6.0 & - & 0.170 & 0.025 & 0.009 & 0.003 & 0.001 \\
			7.0 & - & 0.439 & 0.046 & 0.014 & 0.004 & 0.001 \\
			8.0 & - & 1.249 & 0.084 & 0.023 & 0.006 & 0.001 \\
			9.0 & - & 0.068 & 0.155 & 0.037 & 0.008 & 0.002 \\
			10. & - & - & 0.297 & 0.058 & 0.011 & 0.003 \\
			11. & - & - & 0.606 & 0.091 & 0.016 & 0.003 \\
			12. & - & - & 1.371 & 0.146 & 0.022 & 0.004 \\
			13. & - & - & 3.460 & 0.239 & 0.029 & 0.006 \\
			14. & - & - & 0.064 & 0.404 & 0.040 & 0.007 \\
			15. & - & - & - & 0.722 & 0.055 & 0.009 \\
			16. & - & - & - & 1.392 & 0.074 & 0.011 \\
			17. & - & - & - & 2.997 & 0.102 & 0.013 \\
			18. & - & - & - & 6.992 & 0.141 & 0.016 \\
			19. & - & - & - & 0.063 & 0.197 & 0.020 \\
			20. & - & - & - & - & 0.280 & 0.024 \\
			21. & - & - & - & - & 0.406 & 0.029 \\
			22. & - & - & - & - & 0.602 & 0.035 \\
			23. & - & - & - & - & 0.925 & 0.042 \\
			24. & - & - & - & - & 1.484 & 0.051 \\
			25. & - & - & - & - & 2.528 & 0.062 \\
			26. & - & - & - & - & 4.676 & 0.075 \\
			27. & - & - & - & - & 9.565 & 0.091 \\
			28. & - & - & - & - & 15.420 & 0.111 \\
			29. & - & - & - & - & 0.063 & 0.136 \\
			30. & - & - & - & - & - & 0.167 \\
			35 & - & - & - & - & - & 0.507 \\
			40 & - & - & - & - & - & 2.048 \\
			45 & - & - & - & - & - & 16.880 \\
			50 & - & - & - & - & - & - \\
			\bottomrule
		\end{tabular}
		
	\end{table}
	
	\subsection{$\bar{\nu}_e + d \rightarrow e^+ + n + n  $}
	\begin{table}[H]
		\centering
		\caption{\label{databar} The differential cross section ($d\sigma/dE_e^\prime$ in units of 
			$10^{-42}$ cm$^2$ MeV$^{-1}$) for $\bar{\nu}_e$ CC interaction is tabulated for various 
			choices of outgoing positron energy $E_e^\prime$ and incoming neutrino energy $E_\nu$. }
		\begin{tabular}{c c c c c c c }
			\toprule
			$E_e^\prime\text{ (MeV)}$ & $E_\nu = $ 5 MeV & 10 MeV & 15 MeV &20 MeV &30 MeV & 
			50 MeV \\
			\midrule
			0.5 & - & - & - & - & - & - \\
			1.0 & 0.021 & 0.001 & 0.000 & 0.000 & 0.000 & 0.000 \\
			2.0 & - & 0.007 & 0.002 & 0.001 & 0.000 & 0.000 \\
			3.0 & - & 0.023 & 0.005 & 0.002 & 0.001 & 0.000 \\
			4.0 & - & 0.070 & 0.011 & 0.005 & 0.002 & 0.001 \\
			5.0 & - & 0.219 & 0.023 & 0.008 & 0.003 & 0.001 \\
			6.0 & - & 0.896 & 0.045 & 0.014 & 0.005 & 0.002 \\
			7.0 & - & - & 0.087 & 0.023 & 0.007 & 0.003 \\
			8.0 & - & - & 0.173 & 0.038 & 0.010 & 0.004 \\
			9.0 & - & - & 0.369 & 0.060 & 0.014 & 0.005 \\
			10. & - & - & 0.907 & 0.097 & 0.019 & 0.006 \\
			11. & - & - & 3.156 & 0.159 & 0.026 & 0.007 \\
			12. & - & - & - & 0.269 & 0.035 & 0.009 \\
			13. & - & - & - & 0.483 & 0.046 & 0.011 \\
			14. & - & - & - & 0.946 & 0.063 & 0.014 \\
			15. & - & - & - & 2.168 & 0.085 & 0.017 \\
			16. & - & - & - & 7.180 & 0.115 & 0.020 \\
			17. & - & - & - & - & 0.158 & 0.024 \\
			18. & - & - & - & - & 0.220 & 0.029 \\
			19. & - & - & - & - & 0.313 & 0.034 \\
			20. & - & - & - & - & 0.457 & 0.041 \\
			21. & - & - & - & - & 0.689 & 0.048 \\
			22. & - & - & - & - & 1.087 & 0.057 \\
			23. & - & - & - & - & 1.826 & 0.067 \\
			24. & - & - & - & - & 3.366 & 0.080 \\
			25. & - & - & - & - & 7.257 & 0.095 \\
			26. & - & - & - & - & 9.398 & 0.113 \\
			27. & - & - & - & - & - & 0.135 \\
			28. & - & - & - & - & - & 0.162 \\
			29. & - & - & - & - & - & 0.196 \\
			30. & - & - & - & - & - & 0.238 \\
			35 & - & - & - & - & - & 0.712 \\
			40 & - & - & - & - & - & 3.187 \\
			45 & - & - & - & - & - & 14.900 \\
			50 & - & - & - & - & - & - \\
			\bottomrule
		\end{tabular}
		
	\end{table}
	
	\section{Estimating the number of target nuclei $N_d$}
	\label{estNd}
	
	In this paper, the event rates are obtained for 100\% deuterated hydrocarbon of the generic 
	form 
	$\text{C}_n \text{D}_{2n}$. The density is assumed to be 1 g/cm$^3$. The molar mass of such a 
	compound is $n\times12.0107 + 2n\times 2.014 \approx 16.04 n$ g/mol. For 1 kton of detector 
	volume, the number of deuteron targets is 
	\begin{equation}
		N_{d} = 2n \times \frac{10^9}{16.04n} \times N_A \approx 7.53 \times 10^{31}
	\end{equation}
	where $N_A = 6.023\times10^{23}$ is the Avogadro's number and the prefactor accounts for 
	the fact that there are $2n$ deuteron target per molecule of $\text{C}_n \text{D}_{2n}$. One 
	can obtain estimates for carbon and electron as follows.
	\begin{align}
		N_{T,C} &= \frac{n}{2n} \times N_{T,D} = 3.76 \times 10^{31} \\
		N_{T,e} &= 6 \times N_{T,C} + N_{T,D} = 3.00 \times 10^{32}
	\end{align}
	One can also consider 100\% deuterated linear alkylbenzene ($\text{C}_6 
	\text{D}_5\text{C}_{n}\text{D}_{2n+1}$) based scintillator. For the alkyl group, $n$ usually lies 
	between 10 and 16. Assuming $n=12$, the number of deuteron targets per kton is 
	\begin{equation}
		N_{d}^{dLAB} = 30 \times \frac{10^9}{276.61} \times N_A \approx 6.53 \times 10^{31}
	\end{equation}
	which is about 15\% smaller and the event rates will be scaled accordingly.

	\section{Neutronization Burst}\label{App:Burst}
	
\begin{figure}[!t]
		\centering
		\includegraphics[width=7cm]{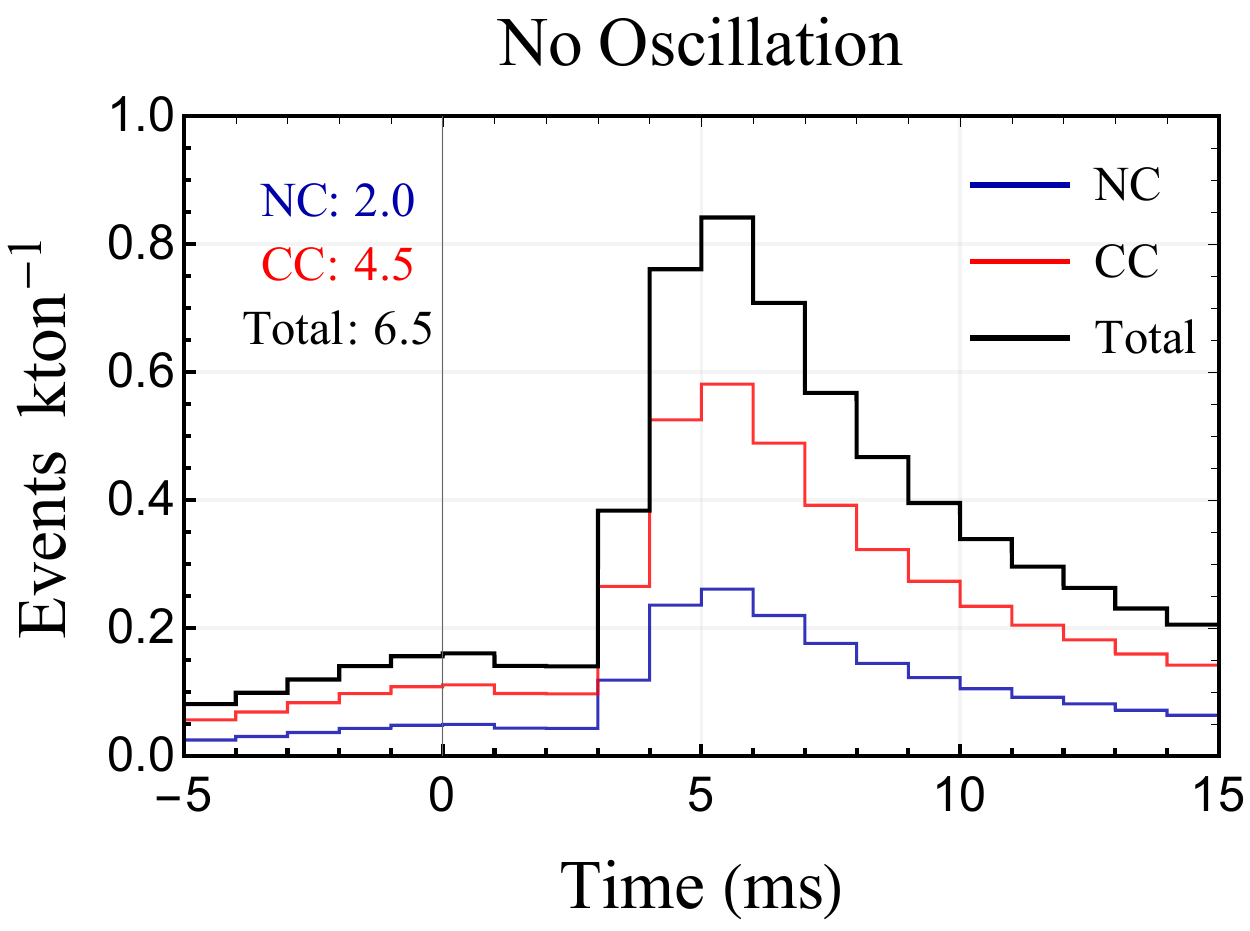}
		\includegraphics[width=7cm]{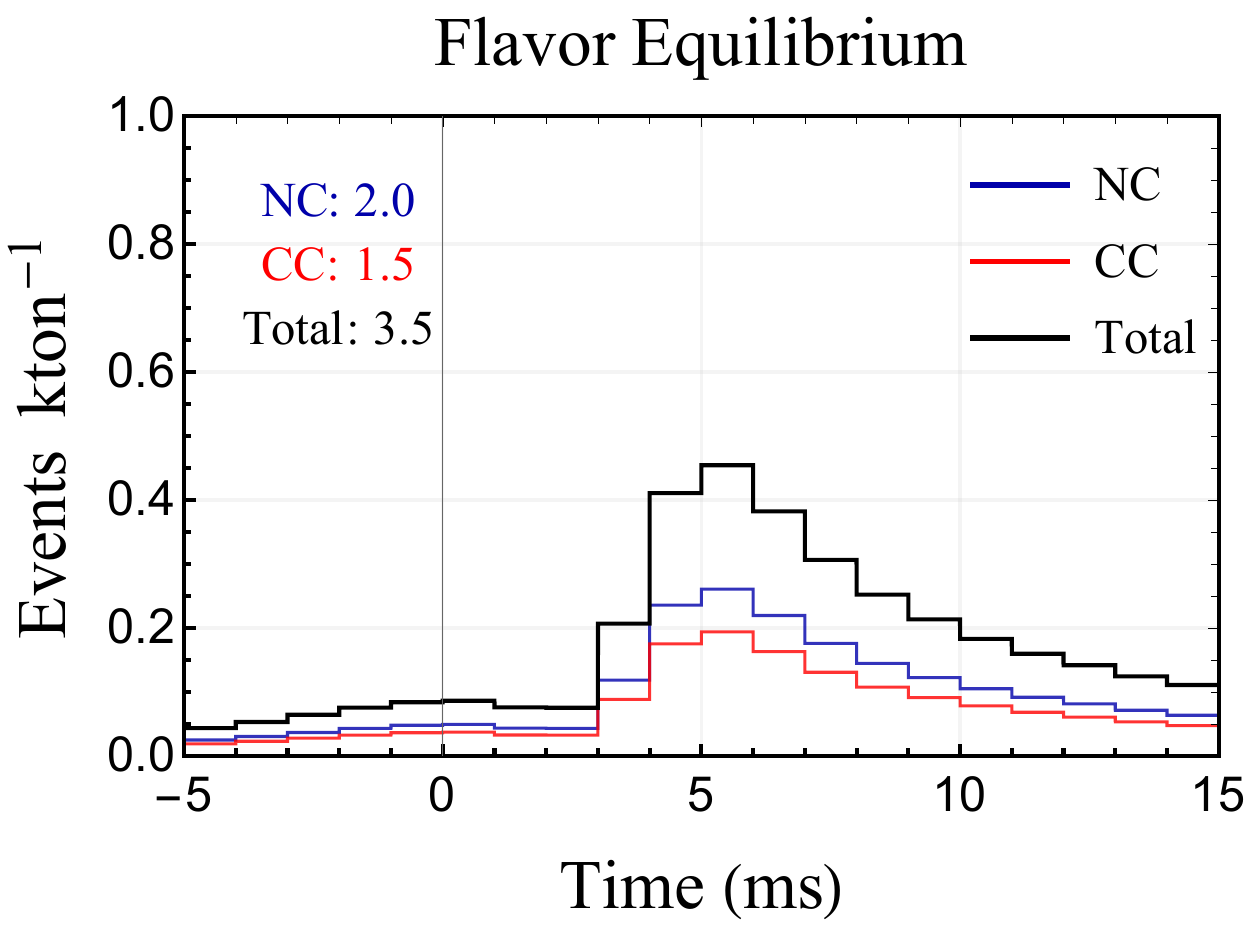}
		\caption{\label{fig:burst} The timing spectrum of events during the neutronization burst for the  
			no oscillation (left) and flavor equilibrium (right) scenario. The aggregate event rate over the 
			course of burst is shown as well.}
	\end{figure}

	A typical neutronization burst lasts about 15-20 ms with peak 
	luminosity of $3.3 - 3.5\times10^{53}$ergs/sec and is nearly independent of progenitor mass 
	and 
	other parameters \cite{Kachelriess:2004ds}. During this burst, not only the luminosity, but also 
	the 
	average energy of the neutrinos changes with time and a simple description, similar to the one 
	we 
	have adopted for fluence, is not sufficient. Deferring details to a future 
	publication, here we show the timing spectra of the events from a neutronization burst. 
	We have adopted the time series of luminosity and average energy from 
	\cite{Kachelriess:2004ds} 
	for a 15 $M_{\odot}$ progenitor, and assume a uniform pinching factor of 3. The events from 
	non-electron flavor neutrinos are negligible. The expected event rate at the detector depends on 
	the neutrino mass hierarchy as well as the flavor conversion parameters. The timing spectrum of 
	events for the two flavor conversion scenarios is 
	shown in Figure \ref{fig:burst}. For the flavor equilibrium case, the NC events are 
	unchanged and the CC events are scaled by a factor of 1/3 as compared to the \emph{no 
		oscillation} scenario. Over the course of the burst, we 
	obtain an aggregate of 6.5 events (2 NC; 4.5 CC) for the no oscillation scenario and 3.5 
	events (2 NC; 1.5 CC) for the flavor equilibrium scenario. 
	
	\bibliographystyle{JHEP}
	\bibliography{references}

\end{document}